\documentclass[11pt]{article}
\usepackage[onehalfspacing]{setspace}
\usepackage[top=1.2in, bottom=1.2in, left=1.0in, right=1.0in]{geometry}

\usepackage{graphicx}
\usepackage{pgfplots}

\usepackage[colorlinks=true,citecolor=blue]{hyperref}

\usepackage{CJK}
\usepackage{CJKnumb}
\usepackage{graphicx}
\usepackage{amsfonts}
\usepackage{array}
\usepackage{amsmath}
\usepackage{amssymb}
\usepackage{url}
\usepackage{fancyhdr}
\usepackage{indentfirst}
\usepackage{enumerate}
\usepackage{titlesec}
\usepackage{amsthm}
\usepackage{mathrsfs}
\usepackage{multirow, multicol}
\usepackage{lscape}
\usepackage{chngpage}
\usepackage{footnote}
\usepackage{algorithm}
\usepackage{algpseudocode}
\usepackage{dsfont,bm}
\usepackage{float}
\usepackage{subfigure}
\usepackage{comment}
\usepackage[toc,page]{appendix}
\usepackage{epstopdf} 
\usepackage[export]{adjustbox} 
\usepackage{rotating}
\usepackage{epstopdf}

\definecolor{darkblue}{rgb}{0.1,0.1,0.6}
\definecolor{darkgreen}{rgb}{0.1,0.5,0.1}

\usepackage[sort&compress,authoryear]{natbib}
\usepackage{cases}

\theoremstyle{definition}

\def\P{\mathbb{P}}
\def\p{\mathbb{P}}
\def\E{\mathbb{E}}

\def\R{\mathbb{R}}
\def\S{\mathcal{S}}

\def\L{\mathcal{L}}

\DeclareMathOperator*{\esssup}{ess\,sup}

\def\TheoremsNumberedThrough{%
	\theoremstyle{TH}%
	\newtheorem{theorem}{Theorem}

	\newtheorem{assumption}{Assumption}
	\theoremstyle{definition}
    \theoremstyle{proof}

	\newtheorem{definition}{Definition}

}

\TheoremsNumberedThrough

\allowdisplaybreaks[4]

\begin{document}

\title{Periodic evaluation of defined-contribution pension fund: A dynamic risk measure approach}
\author{Wanting He\thanks{\rm Department of Statistics and Actuarial Science, The University of Hong Kong, Pokfulam, Hong Kong.  Email: u3006949@connect.hku.hk.} 
\and  Wenyuan Li\thanks{\rm Corresponding author. Department of Statistics and Actuarial Science, The University of Hong Kong, Pokfulam, Hong Kong.  Email: wylsaas@hku.hk.}  
\and Yunran Wei\thanks{\rm School of Mathematics and Statistics, Carleton University, Ottawa, Canada. Email: yunran.wei@carleton.ca.}}
 \maketitle

\begin{center}
  {\bf Abstract}
  \end{center}
  This paper introduces an innovative framework for the periodic evaluation of defined-contribution pension funds. The performance of the pension fund is evaluated not only at retirement, but also within the interim periods. In contrast to the traditional literature, we set the dynamic risk measure as the criterion and manage the tail risk of the pension fund dynamically. To effectively interact with the stochastic environment, a model-free reinforcement learning algorithm is proposed to search for optimal investment and insurance strategies. Using U.S. data, we calibrate pension members' mortality rates and enhance mortality projections through a Lee-Carter model. Our numerical results indicate that periodic evaluations lead to more risk-averse strategies, while mortality improvements encourage more risk-seeking behaviors. 

\vspace{0.5cm}


\section{Introduction}
Due to the aging population, pension management has raised great public concern in recent years. Generally speaking, there are two types of pension schemes: defined-benefit (DB) and defined-contribution (DC) schemes. A DB pension plan is a kind of pension plan in which the benefit is determined by the pension members' salary histories, service years, and retirement age. In contrast, a DC pension plan predetermines the members' contributions during the accumulation period and distributes the retirement benefit based on the investment earnings. Various reasons, such as the increase in longevity and the decrease in interest rates, have raised DB liabilities, prompting a global shift to DC schemes.

The existing literature formulates DC pension management as a portfolio optimization problem during the accumulation period. In this framework, pension members allocate their wealth among various assets (e.g., stocks and bonds) to maximize their investment returns at retirement. Popular criteria can be divided into two categories: (i) utility functions, such as constant relative risk aversion (CRRA), constant absolute risk aversion (CARA), and mean-variance utilities, which quantify individuals' satisfaction; and (ii) disutility functions, such as quadratic loss and risk measures, which capture individuals' dissatisfaction.  The literature on utility-based approaches is extensive, with notable contributions from \cite{boulier2001optimal}, \cite{han2012optimal}, \cite{yao2013markowitz}, \cite{dong2020optimal}, \cite{baltas2022optimal}, \cite{guan2023optimal}, and \cite{pagnoncelli2024solving}. In contrast, research focusing on disutility functions remains limited. Specifically, \cite{haberman2002optimal} minimize quadratic loss between the DC pension fund and a target value via dynamic programming, while \cite{van2010dynamic} computes entropy risk measures without optimization. In general, most literature assumes that DC pension members only evaluate their investment performance at retirement, neglecting pension members' myopic preferences. Moreover, the units of utility and quadratic loss are not currency units (such as dollars), which are hard to measure in practice. These research gaps motivate us to propose a dynamic risk measure approach to study periodic evaluations of the DC pension fund.

A dynamic risk measure is a sequence of conditional risk measures adapted to the filtration. \cite{riedel2004dynamic} pioneers the extension of the coherent risk measure to the dynamic framework with consistency axioms. \cite{ruszczynski2010risk} applies the dynamic risk measure to Markov control processes, deriving dynamic programming equations (DPEs) for finite and infinite horizons. Recent work by \cite{Seb2024} design the RL algorithm to solve time-consistent optimization problems under dynamic risk measures. Periodic evaluation, rooted in prospect theory \cite[see][]{arkes2008reference, shi2015dynamic, tse2023portfolio}, measures rewards as intertemporal wealth changes. This evaluation aligns with pension practices, such as annual performance reviews, rather than a single assessment at retirement. It is natural to employ the dynamic risk measure to study the periodic evaluation of a portfolio, as the difference between the initial and terminal fund values can serve as the random cost in each period.

This paper investigates periodic evaluations of the DC pension account.  A DC pension member dynamically allocates their wealth in risky and risk-free assets and purchases life insurance to hedge their pre-retirement death risk. When a pension member passes away, the beneficiary receives a death benefit that is composed of the pension account and a life insurance payout. A dynamic convex risk measure is employed to characterize the optimal risk-sharing arrangement between the pension member and the beneficiary. In each period, the total risk is modeled as a weighted sum of the pension member's risk associated with investment performance and the beneficiary's risk related to the death benefit. If the pension member survives to the next period, a new risk-sharing arrangement is initiated between the pension member and the beneficiary. 

We formulate this dynamic risk-sharing scheme through the time consistency of the dynamic risk measure. By dual representation, a dynamic programming equation is derived for the underlying optimization problem. We propose a model-free reinforcement learning (RL) algorithm to compute the optimal value function and strategies. The pension member not only optimizes their trading and insurance strategies based on the information available, but also explores the stochastic environment. Specifically, an actor-critic framework employing the M-bootstrap method is used to learn the optimal policy and value function efficiently. This approach combines the advantages of policy-based and value-based methods, enabling iterative policy updates and value estimations. For RL literature, we refer to \cite{jia2022policya}, \cite{jia2022policyb}, and \cite{Seb2024}. 

Our numerical study examines the impacts of mortality improvement and periodic evaluation on optimal investment and insurance strategies. We calibrate a Lee-Carter (LC) model with the U.S. mortality data throughout the pandemic period and project the future mortality rate to manage the pension members' pre-retirement risk. When mortality improvement is incorporated, pension members become more confident in their longevity, leading to increased investment in risky assets and a reduced demand for life insurance. Consequently, the distribution of terminal wealth shifts to the right for those accounting for mortality improvements. Additionally, we examine the impact of periodic evaluation. We find that non-periodic pension members tend to invest more heavily in risky assets and purchase less life insurance, indicating a greater willingness to take risks over longer evaluation periods. As a result, the terminal wealth distribution for non-periodic pension members is shifted to the right relative to that of periodic evaluators.

Our paper contributes to the existing literature by first managing the tail risk of DC pension members' investment dynamically. Our contributions are summarized as follows: (i) We develop a novel dynamic risk-sharing framework for pension fund management that jointly optimizes investment and insurance decisions under periodic evaluation; (ii) We adopt an efficient actor-critic algorithm with provable convergence and adapted policy gradients to interact with the stochastic environment; (iii) We calibrate a Lee-Carter model with the real mortality data and illustrate pension members with mortality improvement tends to be more risk-seeking, investing more in the stock and reducing their demand for life insurance; (iv) We find that periodic evaluations make pension members more risk averse, stabilizing their investment strategy and increasing their life insurance demand.

The remainder of the paper proceeds as follows. Section \ref{sec_rm} introduces the theoretical foundations of dynamic risk measures. Section \ref{sec_model} proposes our DC pension model within a periodic evaluation framework. Section \ref{sec_alg} details the actor-critic algorithm and the derivations of the policy gradient. Section \ref{sec_mortality_compare} discusses the modeling and calibration of the mortality rate using an adapted LC model. Section \ref{sec_num} presents numerical results to illustrate the effects of mortality improvement and evaluation frequency. Finally, Section \ref{sec_conclusion} concludes the paper.

\section{Risk Measures}\label{sec_rm}
This section reviews static and dynamic risk measures, following the framework established in \cite{follmer2011stochastic} and \cite{ruszczynski2010risk}. Consider a filtered probability space $(\Omega,\mathcal{F},(\mathcal{F}_t)_{t\in\mathcal{T}},\mathbb{P})$ with time horizon $\mathcal{T} = \{0,1,\ldots,T\}$, $\mathcal{F}_0 = \{\emptyset, \Omega\}$, and $\mathcal{F}_T = \mathcal{F}$. Let $L^p := L^p(\Omega,\mathcal{F},\mathbb{P})$ denote the $p$-integrable space of random variables (\textit{i.e.,} space with finite $p$-th moment) for $p \geq 1$, and $L^q$ be the $q$-integrable dual space where $1/p + 1/q = 1$. We begin by defining static risk measures.


\begin{definition}[Convex \& Coherent Risk Measure]\label{def:static-risk}
    Let $X,Y \in L^p$, a functional $\rho: L^p \to \mathbb{R}$ is a \textbf{convex risk measure} if it satisfies
    
    1. Cash Invariance: $\rho(X + c) = \rho(X) + c$ for $c \in \mathbb{R}$;
    
    2. Monotonicity: $X \leq Y \implies \rho(X) \leq \rho(Y)$; 
    
    3. Convexity: $\rho(\lambda X + (1-\lambda)Y) \leq \lambda\rho(X) + (1-\lambda)\rho(Y)$ for $\lambda \in [0,1]$.

    \noindent In addition, if $\rho$ preserves
    
    4. Positive Homogeneity: $\rho(cX) = c\rho(X)$ for $c > 0$,

    \noindent then it is called \textbf{coherent risk measure}. 
\end{definition}

To induce the DPE in section \ref{sec:dynamic_prog}, we present the representation theorem (\cite{Seb2024} and \cite{Shapiro2021}) for risk measures. For $Z\in L^p$ and $\xi\in L^q$, we define $\E^\xi[Z] := \sum_{\omega} Z(\omega) \xi(\omega) \p(\omega)$. 

\begin{theorem}\label{rep_thm}(Representation theorem) 
    For a proper (\textit{i.e.}, $\rho(Z)>-\infty$ and its domain is nonempty), lower semi-continuous (i.e., $\rho(W)\le \liminf_{Z\rightarrow W}\rho(Z)$), convex risk measure $\rho$, the dual representation takes the form 
    $$\rho(Z) = \sup_{\xi \in \mathcal U(\p)}\{\E^{\xi}[Z] - \rho^*(\xi)\},$$
    where $\mathcal U(\p) \subset \left\{\xi\in L^q: \sum_{\omega} \xi(\omega)\p(\omega) = 1, \xi \ge 0\right\}$ is the risk envelope and $\rho^*(\xi)$ is the conjugate\footnote{See Definition 4.5-4.6 in \cite{Seb2024} for the definitions of conjugate and biconjugate.} defined as $\rho^*(\xi) := \sup_{X \in L^p} \left(\mathbb{E}^\xi[X] - \rho(X)\right)$. 
    
    Furthermore, if $\rho$ is a coherent risk measure, then the dual representation simplifies to
     $$\rho(Z) = \sup_{\xi \in \mathcal U(\p)}\{\E^{\xi}[Z]\}.$$
\end{theorem}
Next, we denote $L^p_t:= L^p(\Omega, \mathcal{F}_t, \p)$ as the subspace of the filtered probability space. The conditional risk measure is then defined as the projection of the future risk-to-go onto the present.

\begin{definition}[Conditional Risk Measure]\label{def:conditional_risk_meausre}
    Let $X,Y\in L^p_t$, a map $\rho_t:  L^p \rightarrow L^p_t$ is called a \textbf{convex conditional risk measure} if it satisfies
    
    1. Conditional cash invariance: $\rho_t(X+X_t)=\rho_t(X) + X_t$ for $X_t\in L^p_t$;
    
    2. Monotonicity: $\rho_t(X)\le \rho_t(Y)$ for $X\le Y$;
    
    3. Normalization\footnote{We assume $\rho(0) = 0$ unless stated otherwise, as defined as ``risk of zero" in \cite{moresco2024uncertainty}.}: $\rho_t(0) = 0$;
    
    4. Conditional convexity: $\rho_t(\lambda X+(1-\lambda)Y) \le \lambda \rho_t(X)+(1-\lambda) \rho_t(Y)$ for $\lambda\in [0,1]$.
    
    \noindent In addition, if $\rho_t$ satisfies 
    
    5. Conditional positive homogeneity: $\rho_t(cX) = c\rho_t(X)$ for $c\in L_t^p$ with $c>0$, 
    
    \noindent then it is said to be a \textbf{conditional coherent risk measure}.
\end{definition}
	
When $t=0$, the conditional risk measure in the dynamic setting reduces to a static risk measure. Conversely, by extending to the multi-period settings, we can define the dynamic risk measure and the time consistency property based on the conditional risk measure (see \cite{ruszczynski2010risk}).

\begin{definition}[Dynamic Risk Measure]
     A \textit{dynamic risk measure}  is a sequence of conditional risk measures $\{\rho_{t,T}\}_{t\in\mathcal{T}}$ within a $T$-period episode such that $\rho_{t,T}: L^p_{t,T} \rightarrow L^p_t$, for $t\in\mathcal{T}:=\{0, 1, \ldots, T\}$, $L^p_{t,T}:=L^p_t \times \cdots\times L^p_T$.
\end{definition} 

 \begin{definition}[Time Consistency]
    A sequence of dynamic risk measures $(\rho_t)_{t=0,1,...,T}$ is called \textit{time-consistent} if for all $X, Y\in L^p_{t_1,T}$ and for all $t_1,t_2 \in\mathcal{T}, 0\leq t_1<t_2\leq T$, the condition 
    $$\rho_{t_2,T}(X_{t_2},\ldots,X_T) \le \rho_{t_2,T}(Y_{t_2},\ldots,Y_T) \text{ and } X_k=Y_k,\forall k=t_1,\ldots,t_2$$ 
    implies that $\rho_{t_1,T}(X) \le \rho_{t_1,T}(Y)$.
\end{definition}

Under time-consistency, a recursive relationship can be defined via the following theorem (Theorem 4.12 of \cite{Seb2024}).

\begin{theorem}[Recursive Formulation]
     Let $\{\rho_{t,T}\}_{t\in \mathcal T}$ be a time-consistent, dynamic risk measure. It can be shown that $\{\rho_{t,T}\}_{t\in\mathcal T}$ admits the following recursive formulation:
     \begin{equation}\label{recursive}
         \rho_{t,T}(Z_t,...,Z_T) = Z_t + \rho_t(Z_{t+1} + \rho_{t+1}(Z_{t+2} + \cdots +\rho_{T-2}(Z_{T-1} + \rho_{T-1}(Z_T))\cdots)).
     \end{equation}
\end{theorem}

\section{Model Setup}\label{sec_model}

\subsection{Problem Formulation}
Consider a DC pension plan with fund value $W_t$, where an individual enrolls at age $x$ at time $0$ and retires at time $T$ (so the retirement age is $x+T$). During accumulation, the pension member contributes a fixed proportion $\gamma$ of stochastic labor income $\{Y_t\}_{t=0}^{T-1}$ annually and allocates their wealth among a zero-coupon bond $P_t$, a stock index $S_t$, and term life insurance. The DC pension account evolves as follows
$$
W_{t+1} = (W_t + \gamma Y_t - I_t-\alpha_t)\left(\frac{P_{t+1}}{P_t}\right) + \alpha_t\left(\frac{S_{t+1}}{S_t}\right),
$$
where $\alpha_t$ denotes the amount of money invested in stock and $I_t$ the insurance premium\footnote{The incorporation of term life insurance aligns with pension regulations (\cite{IRS2020,li2024optimal}).}. The asset processes (\textit{i.e.}, $ S_t$ and $ P_t$) can be historical data or simulated samples, utilizing our model-free framework. 

We introduce three constraints to fulfill the regulatory requirements.

1. \textbf{Purchasing Constraint:} $I_t \leq W_t + \gamma Y_t$,

2. \textbf{No-Leverage Constraint:} $I_t\geq 0, \alpha_t \geq 0, I_t + \alpha_t \leq X_t + \gamma Y_t$,

3. \textbf{Solvency Constraint:} $W_t\geq 0, Y_t\geq 0$ for all $t$.

We consider a Markov Decision Process (MDP), defined by the tuple $(\mathcal{S}, \mathcal{A}, \mathbf{c}, \P)$, where the finite horizon is given by $\mathcal{T}:= \{ 0, 1, \ldots, T \}$, representing the sequence of $T$ periods within an episode. Specifically, at each time $t\in \mathcal{T}\backslash\{T\}$, the state is described by the vector $s_t:=(t, W_t, Y_t, P_t, S_t)\in\S$, and the action is given by $a_t:=(\alpha_t, I_t)\in\mathcal{A}_t\subset{\mathcal{A}}$, where the $\mathcal{F}_t$-measurable action space is defined as $\mathcal{A}_t:=\left\{(\alpha_t, I_t)\mid \alpha_t \geq 0, I_t\geq 0, \alpha_t+I_t\leq W_t +\gamma Y_t\right\}$. Furthermore, the dynamic risks are captured by the $\mathcal{F}_{t+1}$-measurable periodic costs $c_t:=(s_t,a_t,s_{t+1}) \in \mathbf{c}$. Lastly, we assume that policy $\pi^{\theta}_t=\pi^{\theta}(a_t \mid s_t)$ is parameterized by $\theta\in\Theta$ and learned via Artificial Neural Network (ANN). In the subsequence, we use superscript $\theta_t$ to indicate that the corresponding variables are dependent on the sequence of policies $\{\pi^\theta_t\}_{t\in \mathcal{T}}$. 

Our primal objective is to minimize the expected dynamic risk between the DC pension member and the beneficiary. Based on the recursive formula (\ref{recursive}), we aim to solve the following T-period risk-sensitive RL problem.
\begin{equation}
\begin{aligned}
    \inf \limits_{\theta} {~}&  q_{x}\rho^B_0 \left(c^{B, \theta}_0\right)  + p_{x} \rho^A_0 \bigg( c^{A, \theta}_0 + e^{-r}q_{x+1}\rho^B_{1}\left( c^{B,\theta}_{1}\right) +  e^{-r} p_{x+1} \rho^A_{1}\bigg( c^{A,\theta}_{1}+\cdots\\
    &+ e^{-r} q_{x+T-1}\rho^B_{T-1}\left( c^{B,\theta}_{T-1}\right)  +  e^{-r} p_{x+T-1}\rho^A_{T-1}\left(c^{A,\theta}_{T-1} +  e^{-r}\rho^A_T(- W^{\theta}_T)\right)\cdots \Big)\bigg) \label{obj}\\
    s.t.{~}  &(\alpha^\theta_t, I^\theta_t) \in \mathcal{A}_t, \qquad\text{ for } t\in\mathcal{T}.
\end{aligned}
\end{equation}

In this formulation, $\rho^A_t$ and $\rho^B_t$ quantify risks for the pension member and the beneficiary, respectively. The pension member enters the DC plan at age $x$ at time $t=0$. $p_{x+t}$ is the probability that the pension member age $(x+t)$ survives to the next year $t+1$, $q_{x+t}$ is the probability that pension member age $(x+t)$ dies within the current year $[t,t+1)$, and $e^{-r}$ discounts forward periodic costs. Furthermore, $W^{\theta}_T$ represents terminal payback at retirement, and $c^{A,\theta}_t, c^{B,\theta}_t$ are periodic costs of the pension member and the beneficiary, respectively. At each period, the total risk is modeled as a weighted sum of two components: the pension member's risk stemming from investment performance and the beneficiary's risk related to the death benefit. If the pension member survives to the subsequent period, a new risk-sharing arrangement is initiated between the pension member and the beneficiary.

Lastly, we formulate periodic and non-periodic evaluations through different random costs. The periodic evaluator faces the following random costs for $t \in \mathcal{T}\setminus\{T\}$
\begin{equation}
    c^{A,\theta}_t := W_t + \gamma Y_t - W^\theta_{t+1}, \quad c^{B,\theta}_t :=  - W^\theta_{t+1} - \dfrac{e^r I^\theta_t}{q_{x+t}},\label{pe_cost}
\end{equation}
where $c^A_t$ captures the periodic wealth loss of the pension member's account, and $c^B_t$ represents the lump-sum payment, consisting of the death benefits and the inherited wealth if the pension member dies within the year $[t,t+1)$.

In contrast, under non-periodic evaluation, the pension member considers only terminal wealth while being indifferent to wealth fluctuations during interim periods. Therefore, the random costs for a non-periodic evaluator are
\begin{equation}
    c^{A,\theta}_t := 0, \quad c^{B,\theta}_t := - W^\theta_{t+1} - \dfrac{e^r I^\theta_t}{q_{x+t}}.\label{np_cost}   
\end{equation}

\subsection{Dynamic Programming Equation}\label{sec:dynamic_prog}

To solve the primal objective (\ref{obj}), we derive the corresponding DPE and transform it into an equivalent minimax optimization problem. We first define the transition probabilities
\begin{equation*}
\P^\theta\left(a, s_{t+1} \mid s_t = s\right) := \P\left(s_{t+1} \mid s, a\right) \pi^\theta\left(a \mid s_t = s\right),
\end{equation*}
and $\xi$-weighted conditional expectation as
\begin{equation*}
    \E_t^{\xi}[Z]:=\sum_{\left(a, s_{t+1}\right)} \xi\left(a, s_{t+1}\right) \P^\theta\left(a, s_{t+1} \mid s_t\right) Z\left(a, s_{t+1}\right), \qquad\text{ for }\forall t \in \mathcal{T}.
\end{equation*}
We further assume that all one-step conditional risk measures $\rho_t$ in $\left\{\rho_{t, T}\right\}_{t \in \mathcal{T}}$ satisfy the assumptions in Theorem \ref{rep_thm}. Before using the dual representation, we first define the risk envelope according to Assumption 5.1 in \cite{Seb2024}.
\clearpage
\begin{definition}[Risk Envelope] \label{riskenvelope}
    For convex risk measures $\rho$, the risk envelope $\mathcal{U}(\P^\theta(\cdot,\cdot|s))$ is defined as
    \begin{align*}
    \mathcal{U}\left(\P^\theta(\cdot, \cdot \mid s)\right)=\Bigg\{\xi: & \sum_{\left(a, s^{\prime}\right)} \xi\left(a, s^{\prime}\right) \P^\theta\left(a, s^{\prime} \mid s\right)=1, \xi \geq 0,\\
    & g_e\left(\xi, \P^\theta\right)=0, \forall e \in \mathcal{E}, f_i\left(\xi, \P^\theta\right) \leq 0, \forall i \in \mathcal{I}\Bigg\}
    \end{align*}
    where $g_e\left(\xi, \P^\theta\right)$ is the equality constraint affine in $\xi$ and $f_i$ is the inequality constraint convex in $\xi$, and $\mathcal{E}$ and $\mathcal{I}$ denote the finite set of equality and inequality constraints, respectively. 
    Lastly, we require that $g_e\left(\xi, p\right)$ and $f_i\left(\xi, p\right)$ are twice differentiable in $p$ with bounded first-order derivative \\
    $\max \left\{\max _{i \in \mathcal{I}}\left|\frac{d f_i(\xi, p)}{d p\left(a, s^{\prime}\right)}\right|, \max _{e \in \mathcal{E}}\left|\frac{d g_e(\xi, p)}{d p\left(a, s^{\prime}\right)}\right|\right\} \leq M$ for all $(a, s^{\prime})\in \mathcal{A}\times\mathcal{S}$.
\end{definition}

With the risk envelope, we can convert the primal objective \eqref{obj} to a minimax optimization problem using the dual representation in Theorem \ref{rep_thm}
\begin{small}
\begin{align}
    \min_\theta \max_{\xi_0 \in \mathcal{U}_0}&\bigg\{  
    q_{x}\E_{0}^{\xi^B_{0}} \left[c_0^{B,\theta}\right] + 
    p_{x}\E_{0}^{\xi^A_{0}} \Big[c_0^{A,\theta} + e^{-r}\max _{\xi_1 \in \mathcal{U}_1}\Big\{q_{x+1}\E_{1}^{\xi^B_{1}} \left[c_1^{\theta,B}\right] + p_{x+1}\E_1^{\xi^A_{1}}\Big[c_1^{\theta,A}+
    \cdots \notag\\
    & + e^{-r}\max _{\xi_{T-1} \in \mathcal{U}_{T-1}}\Big\{ q_{x+T-1}\E_{T-1}^{\xi^B_{T-1}}\left[c_{T-1}^{\theta,B}\right] + p_{x+T-1}\E_{T-1}^{\xi^A_{T-1}}\Big[c_{T-1}^{\theta,A} + e^{-r} \max_{\xi_{T} \in \mathcal{U}_{T}} \left\{ \E^{\xi^A_{T}}_{T,s}\left[-W^{\theta}_{T}\right] \right. \notag\\
    & \left.-\rho_{T}^{A,\ast}\left(\xi^A_{T}\right)\right\}\Big] -q_{x+T-1}\rho_{T-1}^{B,\ast}\left(\xi^B_{T-1}\right)-p_{x+T-1}\rho_{T-1}^{A,\ast}\left(\xi^A_{T-1}\right)\Big\}  \cdots\Big] \notag\\
    &-q_{x+1}\rho_1^{B,\ast}\left(\xi^B_1\right)-p_{x+1}\rho_1^{A,\ast}\left(\xi^A_1\right)\Big\}\Big] -q_x \rho_0^{B,\ast}\left(\xi^B_0\right)-p_x \rho_0^{A,\ast}\left(\xi^A_0\right)\bigg\}, \label{minimax}
\end{align}
\end{small}
\noindent where $\xi_t:= (\xi^A_t, \xi^B_t)$ represents the pair of continuous linear functionals in the dual space $Z^\ast$. In the rest of this paper, we use $\mathcal{U}_t:=\mathcal{U}\left(\P^\theta\left(\cdot, \cdot \mid s_t=s_t\right)\right)$ to denote the risk envelope of $\xi_t$ defined in Definition \ref{riskenvelope}. 

We can also apply dual representation to derive the DPE. Back to the primal objective \eqref{obj}, we define the value function as the expected dynamic risk for both the pension member and the beneficiary
\begin{equation*}
    \begin{aligned}  
        V_t(s;\theta) := & q_{x+t}\rho^B_t(c^{B,\theta}_t | s_t=s) + p_{x+t}\rho^A_t\bigg(c^{A,\theta}_t + e^{-r}q_{x+t+1}\rho^B_{t+1}(c^{B,\theta}_{t+1}) +  e^{-r}p_{x+t+1}\rho^A_{t+1}\Big(c^{A,\theta}_{t+1} + \cdots \\
        & + e^{-r}q_{x+T-1}\rho^{B,\theta}_{x+T-1}(c^{B,\theta}_{T-1}) +  e^{-r}p_{x+T-1}\rho^{A,\theta}_{T-1}\big(c^{A,\theta}_{T-1} +  e^{-r}\rho^{A}_T(-W^{\theta}_T)\big)\cdots \Big)\Big| s_t=s\bigg). 
    \end{aligned}
\end{equation*}
For any state $s \in \mathcal{S}$ and time $t \in \mathcal{T}$, the corresponding DPE under policy $\pi^\theta$ is
\begin{equation}
    \begin{aligned} 
        V_t(s;\theta) & = q_{x+t} \rho^B_t\left( W^\theta_t - W^\theta_{t+1} - \dfrac{I^\theta_t(1+r)}{q_{x+t}} \right) + p_{x+t} \rho^A_t\left( W^\theta_t - W^\theta_{t+1} + e^{-r}V_{t+1} \right) \\
        & = q_{x+t} \rho^B_t\left( c^{B,\theta}_t \right) + p_{x+t} \rho^A_t\left( c^{A,\theta}_t + e^{-r}V_{t+1} \right) \qquad \text{for } t \in\mathcal{T}\backslash\{T\},\label{dpe1}\\
        V_{T}(s;\theta) & = \rho^A_{T}\left(- W^{\theta}_{T} \right).
    \end{aligned}
\end{equation}
Applying the dual representation from Theorem~\ref{rep_thm}, we obtain the equivalent minimax problem for \eqref{dpe1} with the following value functions,
\begin{equation}
    \begin{aligned}
        V_t(s;\theta) & = \max_{\xi_t \in \mathcal{U}_t} \bigg\{ q_{x+t}\E^{\xi^B_t}_{t,s}\left[c^{B,\theta}_t\right] + p_{x+t}\E^{\xi^A_t}_{t,s}\left[c^{A,\theta}_t + e^{-r}V_{t+1}(s^\theta_{t+1};\theta)\right]\\
        & \qquad \qquad -q_{x+t}\rho_t^{B,\ast}\left(\xi^B_t\right)-p_{x+t}\rho_t^{A,\ast}\left(\xi^A_t\right) \bigg\},\quad \text{for } t \in\mathcal{T}\backslash\{T\} \label{dpe2} \\
        V_{T}(s;\theta) & = \max_{\xi_{T} \in \mathcal{U}_{T}} \bigg\{ \E^{\xi^A_{T}}_{T,s}\left[-W^{\theta}_{T}\right]-\rho_{T}^{A,\ast}\left(\xi^A_{T}\right)\bigg\} ,
    \end{aligned}
\end{equation}
where $\E^\xi_{t,s}[\cdot] = \E^\xi[\cdot\mid s_t=s]$. This backward recursion enables periodic risk evaluation throughout the pension member's working period. The subsequent section implements an actor-critic algorithm to solve for optimal strategies numerically.

\section{Algorithm}\label{sec_alg}

We develop an actor-critic algorithm to solve the minimax problem (\ref{minimax}), optimizing the value function and strategies iteratively. This approach incorporates an M-bootstrap nested simulation described in \cite{Seb2024} to efficiently estimate conditional risk measures and handle the bivariate action space of investment and insurance decisions.

\begin{figure}[H]
    \centering
    \includegraphics[width=3.0in]{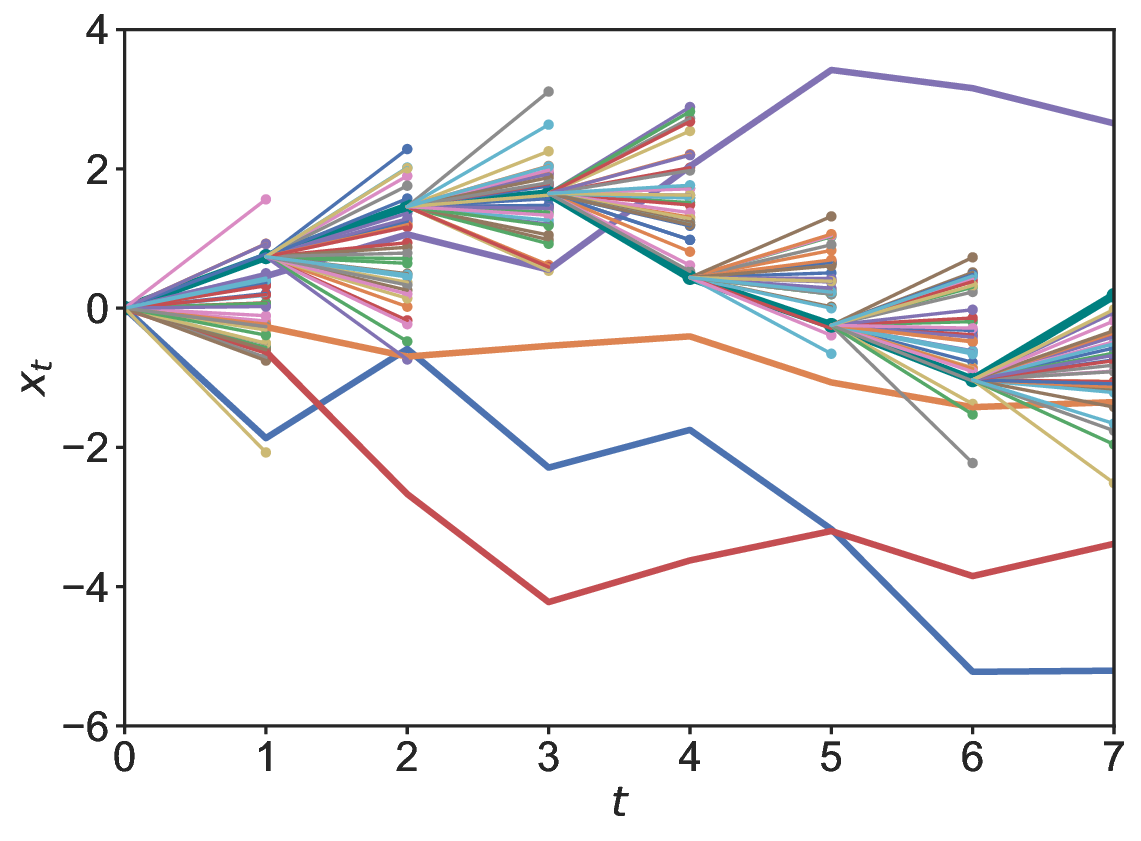}
    \caption{Nested simulation with (Outer) trajectories and (Inner) transitions.}
    \label{fig:bootstrap}
\end{figure}

As shown in Algorithm \ref{alg1}, the actor-critic algorithm alternates between the critic phase (optimizing the value function $V^\phi$ when fixing the policy $\pi^\theta$) and the actor phase (improving the policy  $\pi^\theta$ when fixing the value function $V^\phi$). This dual structure effectively reduces variance in gradient estimates while maintaining exploration capabilities \cite[]{grondman2012survey}. Figure \ref{fig:bootstrap} illustrates our nested simulation approach, where each state evaluation uses $M$ inner transitions sampled from $N$ outer trajectories.  Concretely, we bootstrap $M$ inner transitions $s_{t+1}$ = $\left(s^{(1)}_{t+1},\ldots, s^{(M)}_{t+1}\right)_{M\times 1}$ for each visited state $s_t$ along a single path among $N$ outer trajectories. To accelerate the algorithm, instead of using all the trajectories, we randomly choose $\widetilde{B}$ trajectories (mini-batch size) out of $N$ outer trajectories and then bootstrap $M$ inner transitions solely on the $\widetilde{B}$ trajectories. 

\begin{algorithm}
\caption{Actor-Critic Optimization}\label{alg1}
\begin{algorithmic}
\State \textbf{Initialize} ANNs for value function $V^\phi$ and policy $\pi^\theta$ with number of outer trajectories $N$, inner transitions $M$, epochs $K$, mini-batch size $\widetilde{B}$
\State \textbf{Set} initial guesses for $V^\phi, \pi^\theta$;
\For{each epoch $\kappa=1$ to $K$}
    \State Simulate trajectories using $\pi^\theta$;
    \State Freeze $\tilde{\pi} = \pi^\theta$
    \State \textit{Critic:} Update $V^\phi$ using $\tilde{\pi}$ from Algorithm $2\left(V^\phi, \tilde{\pi}, N, M, K, \widetilde{B}\right)$;
    \State Freeze $\tilde{V}=V^\phi$;
    \State \textit{Actor:} Update $\pi^\theta$ using $\tilde{V}$ from Algorithm $3\left(\tilde{V}, \pi^\theta, N, M, K, \widetilde{B}\right)$;
    \State Store results;
\EndFor
\State \textbf{Return} optimal policy $\pi^{\theta}\approx \pi^{\theta^*}$ and value function $V^\phi\approx V\left(s ; \theta^*\right)$
\end{algorithmic}
\end{algorithm}

Having established the actor-critic framework, we present the detailed estimation of the value function (critic) in Section \ref{subsec_critic} and the update of the policy (actor) in Section \ref{subsec_actor}.

\subsection{Value Function Estimation} \label{subsec_critic}

The critic phase estimates the value function $V^\phi$ by solving the DPEs \eqref{dpe2} through a nested simulation approach. For each state $s_t$, we generate $M$ inner transitions$(s_t, a^{(m)}_t, s^{(m)}_{t+1}, c^{(m)}_t)$ to approximate one-step conditional risk measures. We conduct the simulations over mini-batches $\widetilde{B}$ and optimize the ANN parameters $\phi$ by minimizing the mean squared error between current estimates $V^\phi$ (an ANN estimation on the left-hand side of equation \eqref{dpe2}) and target values $V$ (computed by the right-hand side of equation \eqref{dpe2}). The Adam optimizer (\citeauthor{kingma2014adam}, \citeyear{kingma2014adam}) updates $\phi$ over several epochs to achieve a robust approximation of the value function. The detailed algorithm to compute the critic part is provided in Algorithm \ref{alg2}.

\begin{algorithm}[h!]
\caption{Value Estimation (Critic)}\label{alg2}
\begin{algorithmic}
\State \textbf{Initialize} Value function $V^\phi$, policy $\pi^{\theta}$, number of outer trajectories $N$, inner transitions $M$, epochs $K_V$, mini-batch size $\widetilde{B}$
\For{each epoch $\kappa=1$ to $K_V$}
    \State Set the gradients of $V^\phi$ to zero;
    \State Sample batch of $\widetilde{B}$ states $\{s_t^{(b)}\}_{t \in \mathcal{T}}$;
    \State Generate $M$ transitions per state $\left(a_t^{(b, m)}, s_{t+1}^{(b, m)}, c_t^{(b, m)}\right)_{t\in\mathcal{T}\backslash\{T\} }$ using $\pi^\theta$;
    \For{each state $s_t^{(b)}$}
        \State Compute current estimates $\hat{v}_t^{(b)}=V_t^\phi\left(s_t^{(b)} ; \theta\right)$
        \If{$t=T$}
            \State Calculate target $v^{(b)}_T$ via risk measure optimization
            \State $$ 
            v_{T}^{(b)} = \max _{\xi^A \in \mathcal{U}\left(\P^\theta\left(\cdot, \cdot\mid s_{T}=s_{T}^{(b)}\right)\right)}\left\{\E_{T, s_{T}^{(b)}}^{\xi^A}\left[-W_{T}^{(b, m)}\right]-\rho_{T}^{A,\ast}(\xi^A)\right\};
            $$
        \Else
            \State Calculate target $v^{(b)}_t$ via risk measure optimization
            \State $$ \begin{aligned}
                &\qquad \qquad v_t^{(b)} =  \max _{\xi^A \in \mathcal{U}\left(\P^\theta\left(\cdot, \cdot \mid s_t=s_t^{(b)}\right)\right)}\left\{\E_{t, s_t^{(b)}}^{\xi^A}\left[c_t^{A,(b, m)}+e^{-r}V_{t+1}^\phi\left(s_{t+1}^{(b, m)}; \theta\right)\right]-\rho_t^{A,\ast}(\xi^A)\right\} p_{x+t} \\
                &\qquad \qquad +\max _{\xi^B \in \mathcal{U}\left(\P^\theta\left(\cdot, \cdot \mid s_t=s_t^{(b)}\right)\right)}\left\{\E_{t, s_t^{(b)}}^{\xi^B}\left[c_t^{B,(b, m)}\right]-\rho_t^{B,\ast}(\xi^B)\right\} q_{x+t};
            \end{aligned}   
            $$
        \EndIf
    \EndFor
    \State Accumulate loss $\ell^{(b)}_t = (\hat{v}_t^{(b)} - v_t^{(b)})^2$;
    \State Update $\phi$ using Adam optimizer;
\EndFor
\State \textbf{Return} estimated value function $V_t^\phi(s ; \theta) \approx V_t(s ; \theta)$
\end{algorithmic}
\end{algorithm}

We now establish the theoretical foundation for our neural network approximation of the value function $V_t$. The following theorem demonstrates that our ANN architecture can approximate the true value function to arbitrary precision.

\begin{theorem}(Approximation of $V$)\label{approxV_thm}
    For any fixed policy $\pi^\theta$ with associated value function $V_t(s; \theta)$ defined in \eqref{dpe1}, and for any desired approximation accuracy $\epsilon^* > 0$, there exists a neural network $V_t^\phi:\mathcal S\rightarrow \R$ such that $\esssup_{s\in\mathcal S} ||V_t(s;\theta) - V_t^\phi(s;\theta)||<\epsilon^*,$ holds uniformly across all time periods $t\in \mathcal T$.
\end{theorem}

The proof is delegated to in Appendix \ref{appendA}, which adapts the universal approximation arguments from Section 6.1 of \cite{Seb2024} to our specific context. This theorem guarantees that our neural network architecture can efficiently approximate the value function across all the state space and time horizons.

\clearpage 

\subsection{Policy Iteration} \label{subsec_actor}

Building upon the value function estimates $V^\phi$ from the critic phase, in the actor phase, we optimize our policy $\pi^\theta$ for the joint investment-insurance strategy $a=(\alpha, I)$ using \textit{policy gradient methods} \cite[]{sutton1999policy}. The update rule follows the standard gradient descent method
$$
\theta \longleftarrow \theta + \eta\nabla_{\theta}V(\cdot ; \theta),$$
where $\eta$ denotes an exponentially decaying learning rate that ensures convergence \cite[see Proposition 3.5 in][]{bertsekas1996neuro}, and $\nabla_{\theta}V(\cdot;\theta)$ denotes the gradient of the value function with respect to policy parameters $\theta$.  

The gradient computation relies on three key components: 

 1. \textbf{Differentiable Policy.} We require the policy to satisfy smoothness conditions in Assumption \ref{assump1}. 
\begin{assumption} \label{assump1}
    Suppose $\log \P^{\theta} (a, s^{\prime}|s)$ is differentiable with $\theta$ when $\log \P^{\theta} (a, s^{\prime}|s)\neq 0$, and \\
    $\nabla_\theta \log\P^{\theta}(a,s^\prime\mid s)\leq M$ is bounded for any $(a,s)\in \mathcal{A} \times \mathcal{S}$.
\end{assumption}
 2. \textbf{Gradient Decomposition.} Through the \textit{likelihood ratio trick} (also called \textit{reparametrization trick}), the gradient $\nabla_{\theta}\log\P^{\theta}$  simplifies to 
\begin{equation}\label{probtrick}
\begin{aligned}
    \nabla_{\theta}\P^{\theta}(a,s^{\prime}|s) &=\P^{\theta}(a,s^{\prime}|s)\nabla_{\theta}\log\P^{\theta}(a,s^{\prime}|s) \\
    &= \P^{\theta}(a,s^{\prime}|s)\nabla_{\theta}\left( \log \P (s^{\prime}|a, s) + \log\pi^{\theta}(a|s)\right) \\
    &=\P^{\theta}(a,s^{\prime}|s)\nabla_{\theta}\log\pi^{\theta}(a|s).
\end{aligned}
\end{equation}

 3. \textbf{Risk-Aware Updates.} The complete gradient of the value function is provided in Theorem \ref{gradV_thm}. The corresponding proof is delegated to the Appendix \ref{appendB}.
\begin{theorem}(Policy Gradient)\label{gradV_thm}
    Let Assumption \ref{assump1} hold for the risk envelope defined in Definition \ref{riskenvelope}. For any state $s\in \S$ and time period $t \in \mathcal{T}$, the gradient of the value function (\textit{i.e.}, the policy gradient) is
    $$
    \begin{aligned}
        &\nabla_\theta V_t(s ; \theta)= \E_t^{\xi^{A,\ast}} \left[\left(c_{t}^{A,\theta}+V_{t+1}\left(s_{t+1}^\theta ; \theta\right)-\lambda_A^\ast\right) \nabla_\theta \log \pi^\theta\left(a_t^\theta \mid s_t=s\right)+\nabla_\theta V_{t+1}\left(s_{t+1}^\theta ; \theta\right)\right] p_{x+t} \\
        & + \E_{t}^{\xi^{B,\ast}}  \left[\left(c_{t}^{B,\theta}-\lambda_B^\ast\right) \nabla_\theta \log \pi^\theta\left(a_{t}^\theta \mid s_{t}=s\right)\right] q_{x+t}  -\nabla_\theta \rho_{t}^{A,\ast}\left(\xi^{A,\ast}\right)p_{x+t} - \nabla_\theta \rho_{t}^{B,\ast}\left(\xi^{B,\ast}\right)q_{x+t} \\
        & -\sum_{e \in \mathcal{E}}\lambda^{\ast, \mathcal{E}}(e)\left( p_{x+t}\nabla_\theta g_e\left(\xi^{A,\ast}, \P^\theta\right) + q_{x+t}\nabla_\theta g_e\left(\xi^{B,\ast}, \P^\theta\right)\right) \\
        & -\sum_{i \in \mathcal{I}}\lambda^{\ast, \mathcal{I}}(i)\left( p_{x+t}\nabla_\theta f_i\left(\xi^{A,\ast}, \P^\theta\right) + q_{x+t}\nabla_\theta f_i\left(\xi^{B,\ast}, \P^\theta\right)\right),
    \end{aligned}
    $$
    where $\left(\bm{\xi}^\ast, \bm{\lambda}^\ast, \lambda^{\ast, \mathcal{E}}, \lambda^{\ast, \mathcal{I}}\right):= \left(\xi^{A,\ast}, \xi^{B,\ast}, \lambda^{A,\ast},\lambda^{B,\ast}, \lambda^{\ast, \mathcal{E}}, \lambda^{\ast, \mathcal{I}}\right)$ is any saddle-point of the Lagrangian function of the dual representations \eqref{dpe2},  respectively.

\end{theorem}

When implementing the algorithm, we assume the explicit form of the risk envelope is known in Definition \ref{riskenvelope}, thereby we can obtain the saddle point $\left(\xi^{A,\ast}, \xi^{B,\ast}, \lambda^{A,\ast},\lambda^{B,\ast}, \lambda^{\ast, \mathcal{E}}, \lambda^{\ast, \mathcal{I}}\right)$ of the Lagrangian of equation \eqref{dpe2} for any given risk measure. For instance, the dual representation of CVaR can be written as 
$$ \text{CVaR}_{\alpha}(Z) = \sup _{\xi \in \mathcal{U}(\P)} \E^\xi[Z], $$
where 
$$\mathcal{U}(\P) = \left\{\xi: \sum_{\omega}\xi(\omega)\P(\omega)=1, \xi\in\left[0,\dfrac{1}{\alpha}\right]\right\}.
$$
Any saddle point $(\xi^\ast,\lambda^\ast)$ satisfies $\xi^\ast = \dfrac{1}{\alpha}$ when the risk exposure $Z > q_{1-\alpha}$, and $\xi^\ast = 0$ otherwise; $\lambda^\ast$ is the $(1-\alpha)$-quantile of $Z$\footnote{The interested reader can find more details in Chapter 5 and Chapter 6 in \cite{Shapiro2021}, where both the analytical derivation and sample average approximation over risk measures are provided.}. The policy iteration algorithm for the actor phase is illustrated in Algorithm \ref{alg3}. In this phase, we optimize $\pi^{\theta}$ while keeping $V^{\phi}$ fixed, treating the parameters $\phi$ as constant. The gradient of the value function, as presented in Theorem \ref{gradV_thm}, serves as the policy gradient and is estimated by averaging over mini-batches of states across multiple periods. In addition, when calculating the policy gradient $\nabla_\theta V_{t}$ using the DPE \eqref{dpe2}, the gradient of the next period $\nabla_\theta V_{t+1} = \nabla_\theta V^{\phi}_{t+1} = 0$ vanishes due to the fixed critic part $V^{\phi}_{t+1}$.

\begin{algorithm}[h!]
\caption{Policy Update (Actor)}\label{alg3} 
\begin{algorithmic}
\State \textbf{Initialize} Value function $V^\phi$, policy $\pi^{\theta}$, number of outer trajectories $N$, inner transitions $M$, epochs $K_{\pi}$, mini-batch size $\widetilde{B}$
\For{each epoch $k=1$ to $K_{\pi}$}
    \State Set the gradients of $\pi^\theta$ to zero ;
    \State Sample batch of $\widetilde{B}$ states $\{s_t^{(b)}\}_{t \in \mathcal{T}}$;
    \State Generate $M$ transitions per state $\left(a_t^{(b, m)}, s_{t+1}^{(b, m)}, c_t^{(b, m)}\right)_{t\in\mathcal{T}\backslash\{T\} }$ using $\pi^\theta$;
    \For{each state-transition pair}
        \State Obtain $\hat{z}_t^{(b, m)}=\nabla_\theta \log \pi^\theta\left(a_t^{(b, m)} \mid s_t^{(b)}\right)$ with the reparametrization trick;
        \State Obtain one-step forward values $\hat{v}_{t+1}^{(b, m)}=V_{t+1}^\phi\left(s_{t+1}^{(b, m)} ; \theta\right)$;
        \State Get a saddle-point $\left(\bm{\xi}^*, \bm{\lambda}^*, \lambda^{*, \mathcal{E}}, \lambda^{*, \mathcal{I}}\right)$ and compute $\bm{\xi}_t^{*,(b, m)}=\bm{\xi}^*\left(a_t^{(b, m)}, s_{t+1}^{(b, m)}\right)$;
        \State Obtain $\hat{\rho}_t^{l,(b)}=\nabla_\theta \rho_t^{l}\left(\xi^{l,*}\right), \hat{g}_{e, t}^{l,(b)}=\nabla_\theta g_e\left(\xi^{l,*}, \P^\theta\right)$, and $\hat{f}_{i, t}^{l,(b)}=\nabla_\theta f_i\left(\xi^{l,*}, \P^\theta\right)$ for
        \State $l\in\{A,B\}$;
        \State Calculate the gradient $\nabla_\theta V_t\left(s_t^{(b)} ; \theta\right)$, 
        $$
        \begin{footnotesize}
        \begin{aligned}
            &\ell_t^{(b)} = \frac{1}{M} \sum_{m=1}^M\Bigg\{ p_{x+t}\Bigg[\xi_t^{A,*,(b, m)}\left(c_t^{A,(b, m)}+e^{-r}\hat{v}_{t+1}^{(b, m)}-\lambda^{A,*}\right) \hat{z}_t^{(b, m)}-\hat{\rho}_t^{A,(b)} \\
            & -\sum_{e \in \mathcal{E}} \lambda^{A,*, \mathcal{E}}(e) \hat{g}_{e, t}^{A,(b)}-\sum_{i \in \mathcal{I}} \lambda^{A,*, \mathcal{I}}(i) \hat{f}_{i, t}^{A,(b)}\Bigg] + q_{x+t}\Bigg[\xi_t^{B,*,(b, m)}\left(c_t^{B,(b, m)}-\lambda^{B,*}\right) \hat{z}_t^{(b, m)}-\hat{\rho}_t^{B,(b)} \\
            & -\sum_{e \in \mathcal{E}} \lambda^{B,*, \mathcal{E}}(e) \hat{g}_{e, t}^{B,(b)}-\sum_{i \in \mathcal{I}} \lambda^{B,*, \mathcal{I}}(i) \hat{f}_{i, t}^{B,(b)}\Bigg] \Bigg\} ;
        \end{aligned}
        \end{footnotesize}
        $$  
    \EndFor
    \State Average gradients over batch and across time, $$\ell=\frac{1}{B T} \sum_{b=1}^B \sum_{t=0}^{T-1} \ell_t^{(b)} ;$$
    \State Update $\theta$ using Adam optimizer;
\EndFor
\State \textbf{Return} updated policy $\pi^\theta$
\end{algorithmic}
\end{algorithm}

\section{Lee-Carter Mortality Projection with Transitory \\Jumps and Renewal Effects}\label{sec_mortality_compare}
To better manage longevity risk, we estimate pension members' mortality rates using a Lee-Carter (LC) model (\cite{lee1992modeling}) that incorporates transitory jumps and renewal processes. This modification addresses limitations inherent in traditional ARIMA and broken-trend stationary models, particularly in capturing mortality shocks following the COVID-19 pandemic. Let $m_{x,t}$ denote the central mortality rate at age $x$ in year $t$. The LC model articulates the logarithm of central death rates through the equation
\begin{equation} \label{lc}
    \ln \left(m_{x,t}\right)=a_x + b_x \kappa_t + \epsilon_{x,t},
\end{equation}
where $a_x$ denotes the temporal average of $\ln(m_{x,t})$, $\kappa_t$ reflects long-term mortality improvements, $b_x$ quantifies the sensitivity of mortality with respect to variations in $\kappa_t$, and $\epsilon_{x,t}$ denotes the error term that captures the age and time effects not reflected in the model. Assuming a uniform distribution of deaths between ages, the probability of death occurrence at age $x$ can be expressed by the central death rate as $q_x = m_x/(1+0.5m_x)$.

\clearpage

The estimation of model parameters proceeds in two phases. The first phase utilizes singular value decomposition (SVD) to estimate $a_x, b_x, \kappa_t$ with identifiability constraints ($\sum_x b_x = 1$ and $ \sum_t\kappa_t = 0$) to ensure the uniqueness of the estimation. Once $a_x$ and $b_x$ are obtained, the parameters $\kappa_t$ are re-estimated in the second phase to align the predicted mortality with observed data. More specific, we assume $\kappa_t$ follows a jump-diffusion process proposed by \cite{ozen2020transitory} \footnote{Recent advances in stochastic mortality modeling have extensively explored adaptations of LC framework, including transitory jumps (\cite{chen2009modeling}), renewal processes (\cite{ozen2020transitory}), pandemic-driven shocks (\cite{boado2022pandemics}), and robust forecasting techniques (\cite{csahin2024quantitative}).}
\begin{equation} \label{kappa_t}
    \kappa_t=\kappa_0+\left(\mu-\frac{1}{2} \sigma^2-\delta \theta\right) t+\sigma Z_t+\sum_{i=1}^{N_t} Y_i,
\end{equation}
where $Z_t$ is a standard Brownian motion capturing continuous mortality fluctuations, $N_t$ is a renewal process modeling shock occurrences, and $Y_i\overset{\mathrm{iid}}{\sim}\operatorname{Exp}(\theta)$ are Exponential jump sizes representing mortality shock magnitudes. In addition, $\theta$ denotes the mean jump size and $\delta$ represents the jump arrival rate via the location-scale method  (see \cite{rigby2005generalized} and \cite{ferguson1962location}). 

\begin{figure}[H]
    \centering
    \subfigure[$a_x$]{
        \includegraphics[width=2.6in]{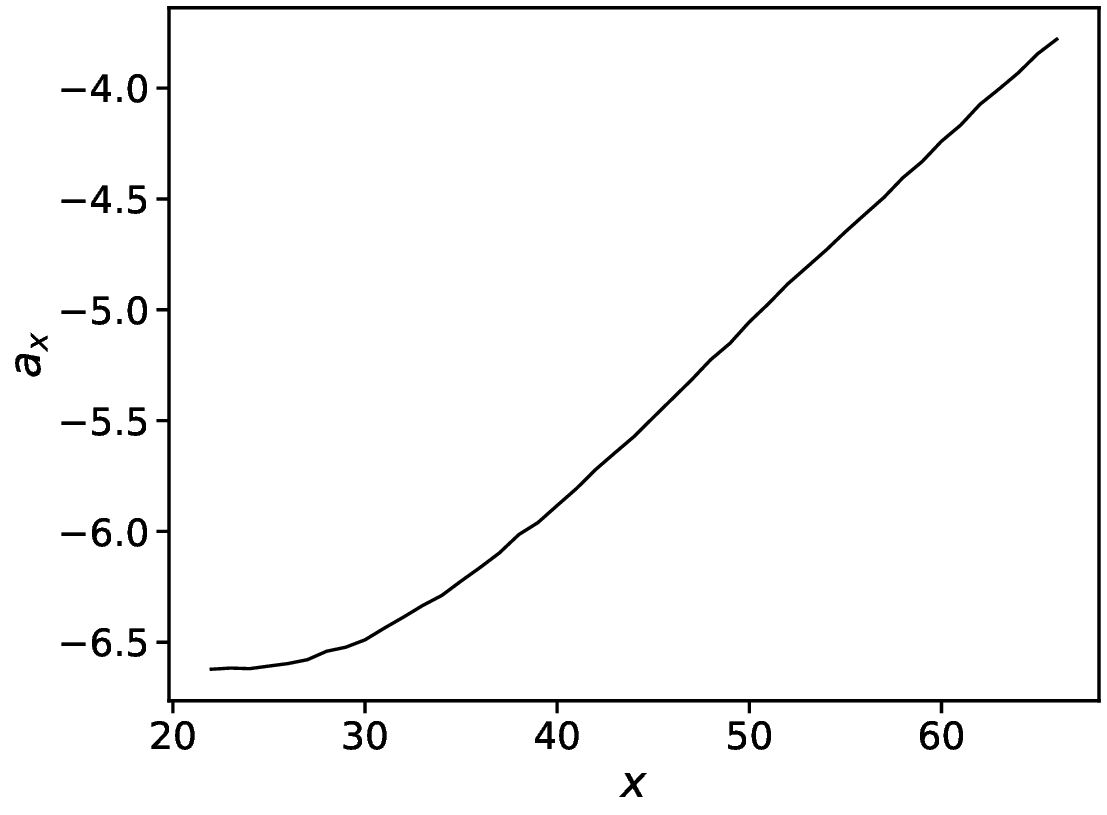}
        \label{fig:lc_ax}
        }    
    \subfigure[$b_x$]{
        \includegraphics[width=2.6in, height=1.92in]{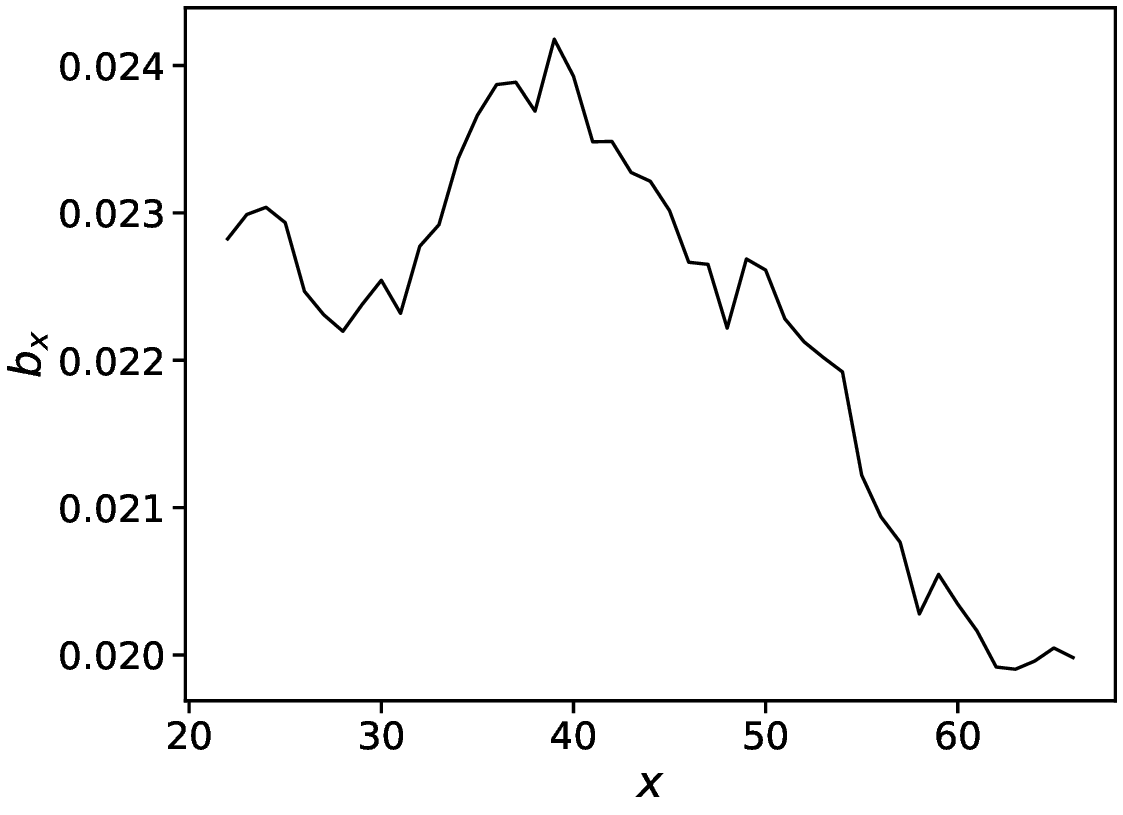}
        \label{fig:lc_bx}
        }
    \label{fig:lc_params}
    \subfigure[$\kappa_t$]{
        \includegraphics[width=2.6in]{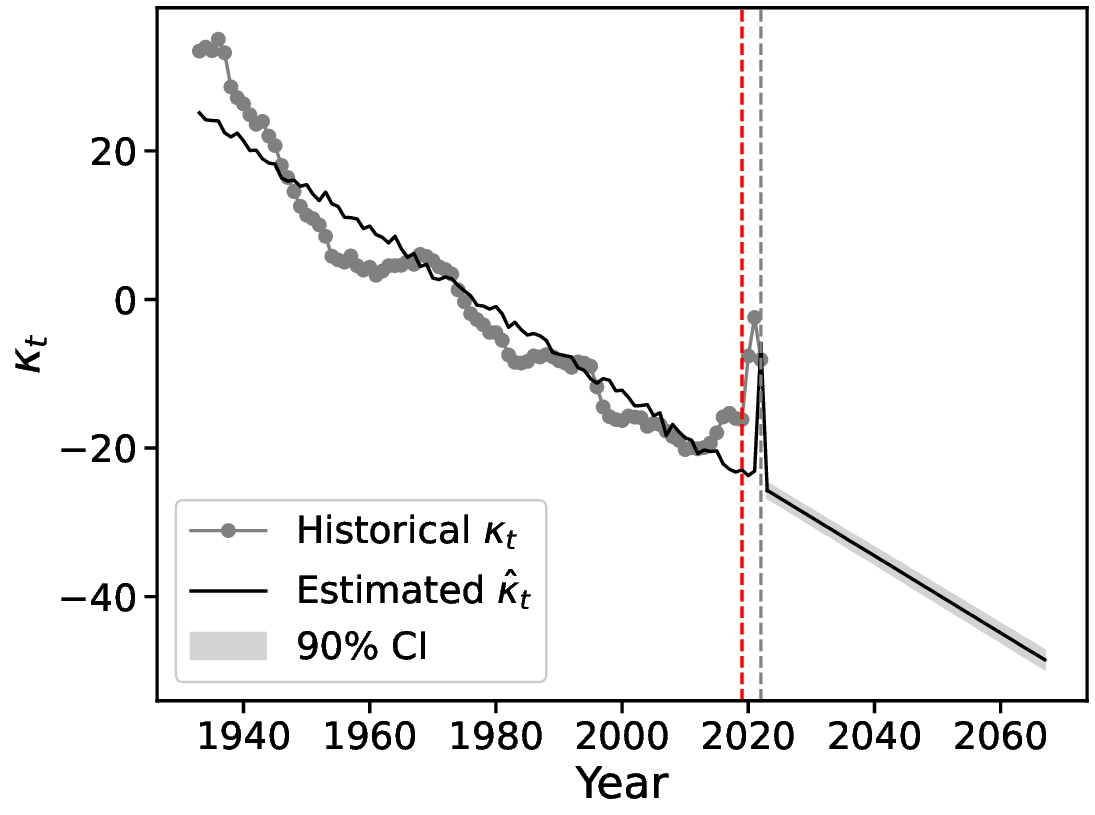}
        \label{fig:lc_kt}
        }    
    \subfigure[Central death rates]{
        \includegraphics[width=2.62in, height=1.95in]{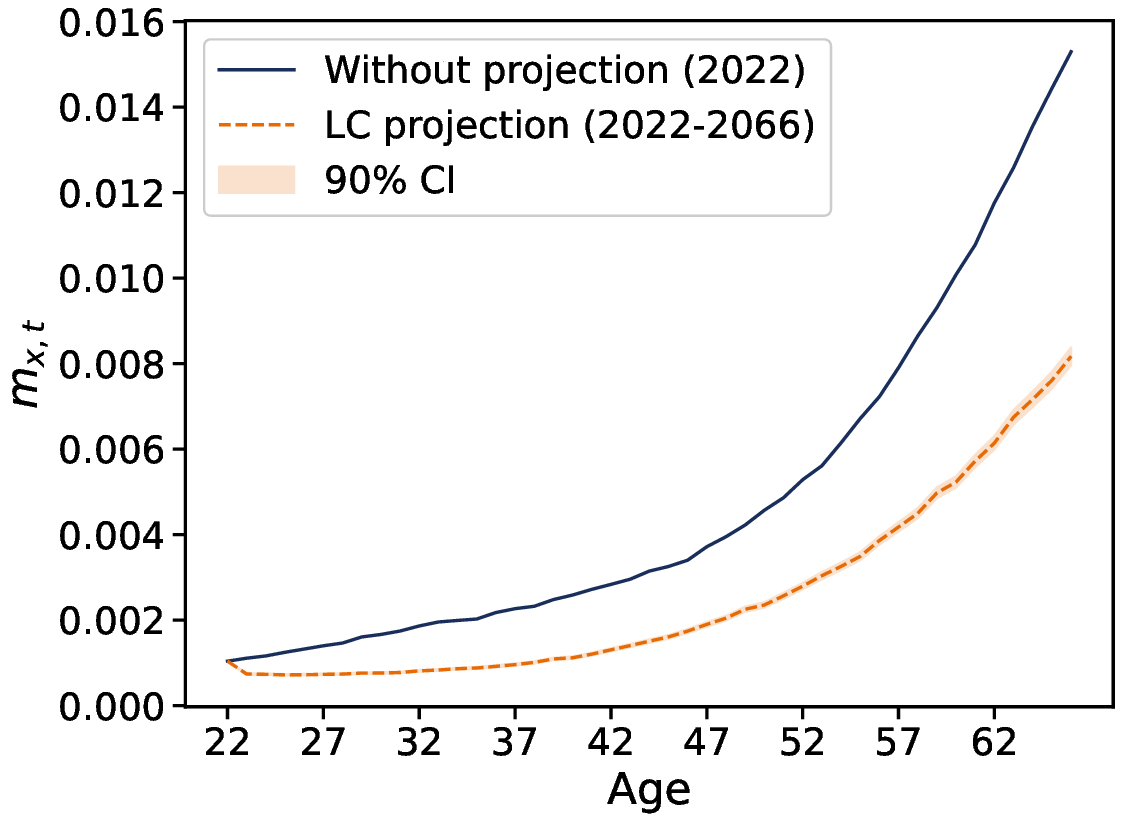}
        \label{fig:lc_compare}
        }
    \caption{The Lee-Carter mortality projection. Figures (a) and (b) show the estimation results for $a_{x}$ and $b_{x}$. Figure (c) is the time-specific parameter $\kappa_t$, with historical data (grey dots, pre-2022), fitted values (black line, pre-2022), and projections (black line, post-2022) with $90\%$ CI (grey area); the red dashed line marks the onset of outlier events (pandemic shock, 2019), and the grey dashed line is the end of historical data (2022). Figure (d) illustrates the comparison over the cohort mortality starting at age 22 in 2022, without projection (blue, 2022) and incorporating the LC projection (orange, 2022-2066) with 90\% CI (orange area).}
\end{figure}

We utilize U.S. mortality data from 1933 to 2022 in \cite{HMD2024} to fit Lee-Carter parameters and forecast $\kappa_t$ (see Figures \ref{fig:lc_ax} to \ref{fig:lc_kt}). Following \cite{chen1993joint}, we detect transition shocks when events exceeding $C=2.0$, with the years 2020, 2021, and 2022 considered as outliers. Additionally, we assume that the inter-arrival times between shocks follow one of three distributions: lognormal, Weibull, or Gamma. The optimal distribution is selected based on the lowest Bayesian Information Criterion (BIC), as outlined in \cite{mcshane2008count}. Using this framework, we project mortality rates for a cohort at age 22 in 2022 over the subsequent 45 years (2022–2066). Figure \ref{fig:lc_compare} contrasts the LC projections (in orange) against static mortality rates (in blue) in 2022, illustrating that pension members have lower central death rates due to the mortality projection. Moreover, the estimated parameters are presented in Table \ref{tab:params}. Our findings indicate that the lognormal distribution provides the best fit for the inter-arrival times between mortality shocks. Overall, the Lee-Carter model effectively captures both long-term mortality trends and transient shocks, offering robust inputs for pension optimization, as demonstrated in Section \ref{sec_num}. 

\begin{table}[ht]
    \caption{Estimation results of $\kappa_t$.}
    \smallskip
    \centering
    \begin{tabular}{>{\centering\arraybackslash}m{2.8cm} 
                >{\centering\arraybackslash}m{3.5cm} 
                >{\centering\arraybackslash}m{3.5cm} 
                >{\centering\arraybackslash}m{3.5cm}}\hline 
        $N_t$'s Model & Lognormal & BIC & $534.8575$ \\ \hline
        Parameter &  $\kappa_0$ & $h^\ast$ & $\sigma$ \\
        Estimate & $2.483138\times 10^1$ & $-5.618974 \times 10^{-1}$ & $5.947313 \times 10^{-1}$  \\
        Parameter &  $\theta$ &  $\delta$ & \\
        Estimate   & $1.155890 \times 10^1$  & $9.999999\times 10^{-1}$ & \\ \hline
        \multicolumn{4}{l}{\scriptsize $\ast$ $h:= \mu-\frac{1}{2} \sigma^2-\delta \theta$ is the drift term parameter in equation (\ref{kappa_t}).}
    \end{tabular}    
    \label{tab:params}
\end{table}

\section{Numerical Illustration}\label{sec_num}
We implement our dynamic risk measure framework using \textit{conditional value-at-risk} (CVaR) with threshold  $\alpha\in(0,1)$, following its demonstrated effectiveness (see \cite{Seb2024}). Unlike traditional expected utility approaches, CVaR specifically targets tail risk management, a critical consideration for pension funds where extreme losses can lead to irreversible consequences. Our analysis examines three key dimensions: investment-insurance strategy optimization, mortality projection impacts, and evaluation frequency effects.  All computations are executed on an NVIDIA RTX A5500 GPU. The implementation code is publicly accessible on GitHub at: \href{https://github.com/WantingHe/DC_Pension-DynamicRiskMeasure-RL}{https://github.com/WantingHe/DC\_Pension-DynamicRiskMeasure-RL} 

\subsection{Market Dynamics}
To illustrate our periodic evaluation framework, we first specify the market dynamics for the DC pension management. We assume that both fund evaluation and strategic adjustments occur on an annual basis. In particular, the dynamic of the risk-free bond $P_t$ is given by
$$
P_{t+1} = P_t e^r,
$$
where the constant $r$ is the interest rate. The dynamic of stock $S_t$ follows a geometric Brownian motion (GBM)
$$
S_{t+1} = S_t \exp(\mu_s + \sigma_s Z_{1,t}),
$$
where $\mu_s$ represents the drift term in the exponential, $\sigma_s$ denotes the volatility of the stock index, and $Z_{1,t}$ is a standard Brownian motion. We also consider a stochastic labor income $Y_t$ evolving with a lognormal growth rate
$$
Y_{t+1} = Y_{t} \exp(\mu_y + \sigma_y Z_{2,t}),
$$
where $\mu_y$ denotes the deterministic growth rate of income. $\sigma_y$ denotes the volatility associated with the income process, and $Z_{2,t}$ is the income's Brownian motion correlated with stock's Brownian motion $Z_{1,t}$ with $\rho_{s,y}=0.3755$\footnote{\cite{munk2010dynamic} report the evidence of the statistically significant correlation coefficient between the stock and the income, using the calibrated data from the Panel Study of Income Dynamics (PSID) survey of the annual income of U.S. individuals and the S\&P 500 index over the period 1951–2003.}, \textit{i.e.}, $\E[Z_{1,t} Z_{2,t}] = \rho_{s,y}$.

In addition to investment options, the pension members are allowed to purchase one-year term life insurance to hedge against mortality risk (see the U.S. regulations \cite{IRS2020}). Let $B_t$ denote the death benefit payable to the beneficiary at time $t+1$ when a pension member dies within the interval $(t, t+1]$. The life insurance is priced under actuarial fairness, with the premium
$$I_t = e^{-r}q_{x+t}B_{t},$$ 
where $q_{x+t}$ represents the probability that an individual aged $x+t$ dies before age $x+t+1$ (dies within the current year) and $e^{-r}$ is the discounting factor.

We consider an individual initiating a DC pension fund at age $x=22$ at time $t=0$. The insured (agent A) and the beneficiary (agent B) adopt CVaR with a commonly selected confidence level of $\alpha_A = \alpha_B = 0.1$. Following the U.S. regulations, the legal retirement age is set at $67$, resulting in a total working time horizon of $T=45$ years. Assuming that the investment-insurance allocation strategy is updated annually, the number of episodes is $N_{\Delta t}=45$. The stock's parameters are chosen as $\mu_s=0.15$, $\sigma_s=0.2$, and the salary process is $\mu_y=0.03$, $\sigma_y=0.05$. Finally, the initial state vector is set as $s_0 = (t_0, w_0, y_0)=(0,5,60)$, where initial wealth $w_0=5$ (in thousand dollars) and initial salary $y_0=60$ (in thousand dollars) at time $t_0=0$. Here, we don't include $P_t$ and $S_t$ in the state variable as $P_{t+1}/P_t$ and $S_{t+1}/S_t$ are irrelevant to state $P_t$ and $S_t$, respectively.

Inspired by \cite{Seb2024}, we assume the distributional policy follows a bivariate normal distribution and use a neural network to learn the parameters of the distribution  
\begin{equation*}
    \pi^{\theta}(a_t|s_t) \sim \text{BiNormal}\left(\begin{bmatrix}\mu_{\alpha}\\ \mu_I\end{bmatrix},\begin{bmatrix} \sigma^2_{\alpha}& \rho \sigma_{\alpha}\sigma_{I}\\ 
    \rho \sigma_{\alpha}\sigma_{I}& \sigma^2_{I} \end{bmatrix}\right),
\end{equation*}
where $\theta$ are the weights and biases of the neural network, $a_t=(\alpha_t, I_t)$ are actions and $s_t=(t,W_t,Y_t)$ are state variables. To improve the convergence, we add bound constraints on the standard deviations that $\sigma_{\alpha}\leq 0.2\mu_{\alpha}$ and $\sigma_{I}\leq 0.2\mu_{I}$. 

For the algorithm architecture, we alternate between actor and critic updates for $K=100$ iterations in the outer loop, using a mini-batch size of $\widetilde{B}=200$. In the inner loops, we repeat $K_V=100$ epochs for the critic estimation and $K_{\pi}=20$ epochs for the actor update. We perform Monte Carlo simulations involving $N=750$ outer trajectories and $M=500$ inner transitions, using a time step of one year. To obtain a good starting point, we run an extra 100 epochs for the critic estimation at initialization. The initial learning rate for the critic part is set to $1\times 10^{-4}$, while the learning rate for the actor is set to $5\times 10^{-7}$, and both exponentially decay by $0.9$ at every epoch. Both actors' and critics' ANNs have four hidden layers with 32 hidden nodes and SiLU activation functions. The output layer of the critic network does not incorporate an activation function, given that the CVaR's range spans the entire real line. For the policy network's output layer, a tanh activation function is used to parameterize the correlation coefficient $\rho \in [0,1]$, while sigmoid activations are applied to other parameters, which are subsequently scaled to satisfy the purchasing and bound constraints.

\subsection{Mortalilty Projection Effect}\label{subsec_period}

\begin{figure}[htbp]
    \centering
    \subfigure[Training loss of value function]{\label{fig:training_loss_V}
        \includegraphics[width=0.4\textwidth]{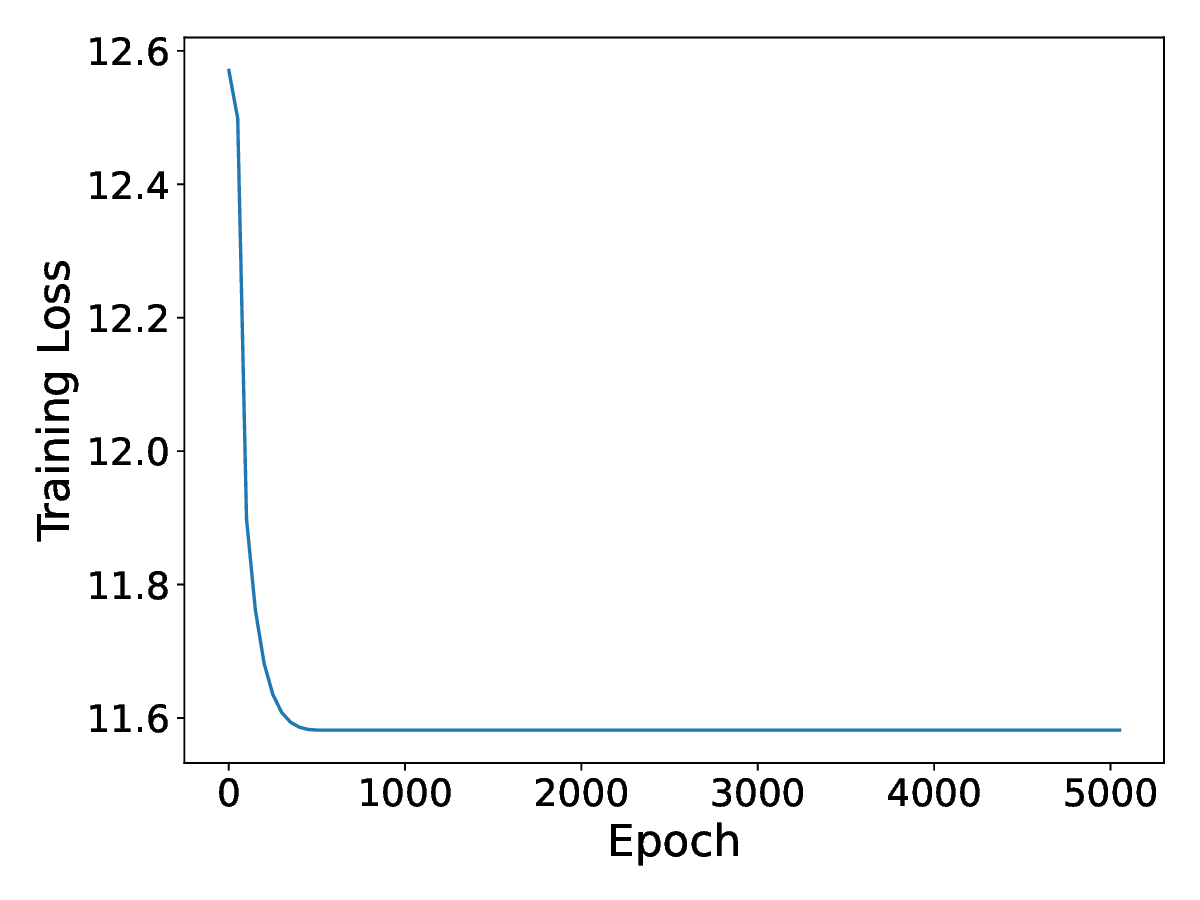}}
    \hspace{1em}
    \subfigure[Training loss of policy iteration]{\label{fig:training_loss_pi}
        \includegraphics[width=0.43\textwidth]{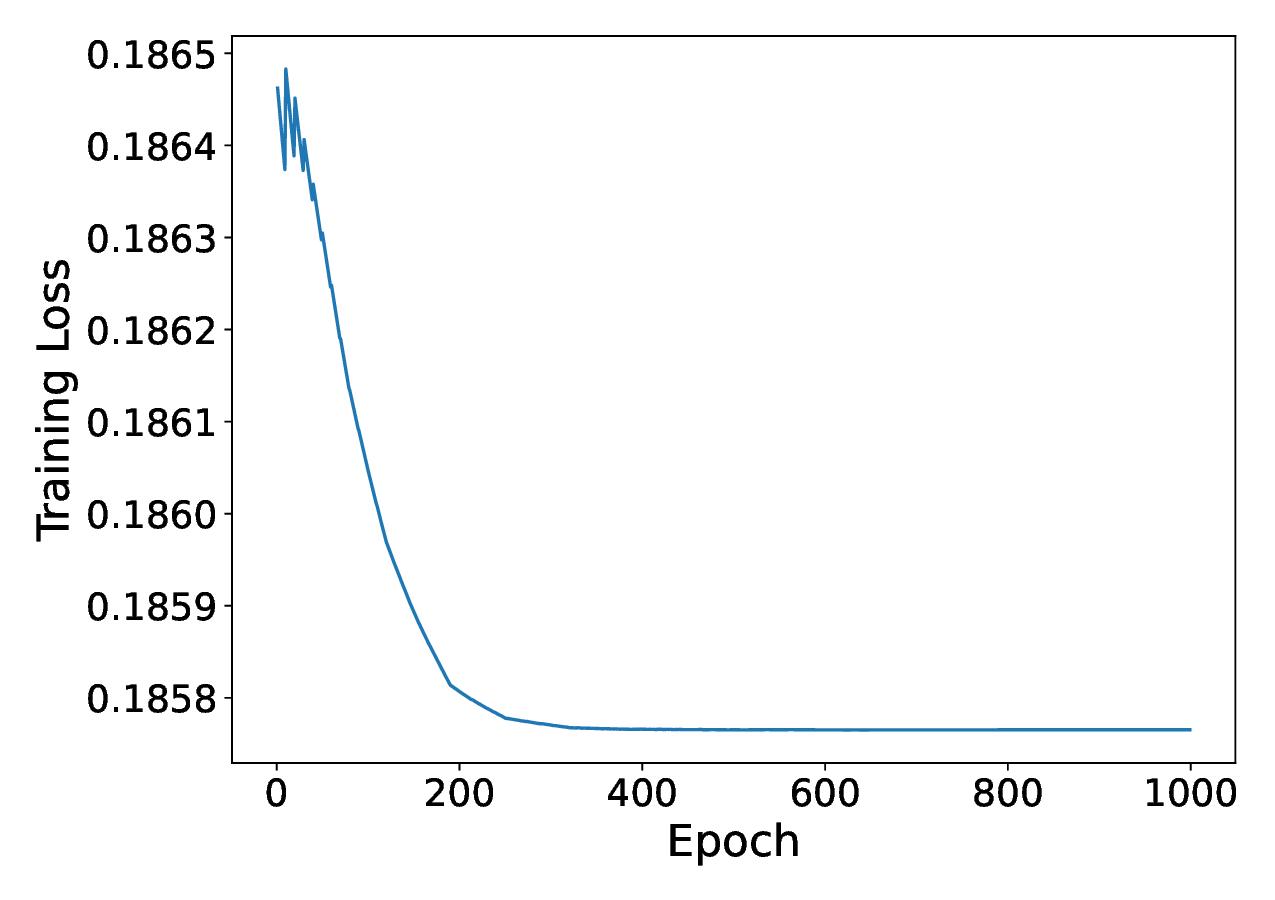}}
    \caption{Training loss of the actor-critic algorithm.}
    \label{fig:training_loss}
\end{figure}

This section illustrates the mortality projection effect on the optimal strategies. Figure \ref{fig:training_loss} shows the loss curves of the value function and the policy iteration. We see that the algorithm is stable and both the value function and policy converge in finite steps.

Table \ref{tab:comparison} shows the numerical results for the distribution of the optimal policy $\pi^{\theta}(\alpha_t, I_t|s_t)$ at selected ages 22, 44, and 66 with and without mortality projection. It reveals a negative correlation between allocations to risky assets and term life insurance. This finding highlights the main differences between these two asset classes: stock is viewed as a risky asset to generate investment earnings, while life insurance is regarded as a protection for future income. For pension members without mortality projection, they reduce allocations to risky investments and increase their life insurance purchases when they age. With the LC projection, the pension member tends to increase their risky investment in the early and middle ages and then decrease towards retirement. This risk-seeking investment comes from the mortality improvement by the LC projection. As a result, the LC projection individuals have a larger wealth compared to those without mortality projection, and their demand for life insurance increases due to the larger wealth. 

\begin{sidewaysfigure}[htbp]
\centering 
\centering
    \subfigure[Periodic evaluation without mortality projection (2022)\label{fig:dynamic}]{\includegraphics[width=\linewidth,height=0.28\textheight,keepaspectratio]{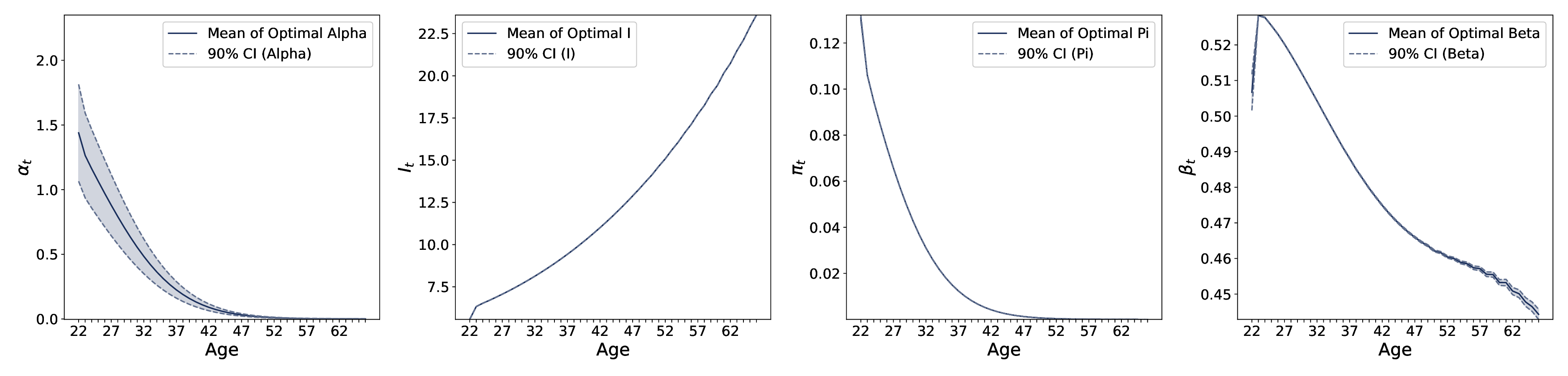}}

\vspace{-0.05in}
\subfigure[Periodic evaluation with LC mortality projection (2022-2066)\label{fig:dynamic_lc}]{
\includegraphics[width=\linewidth,height=0.28\textheight,keepaspectratio]{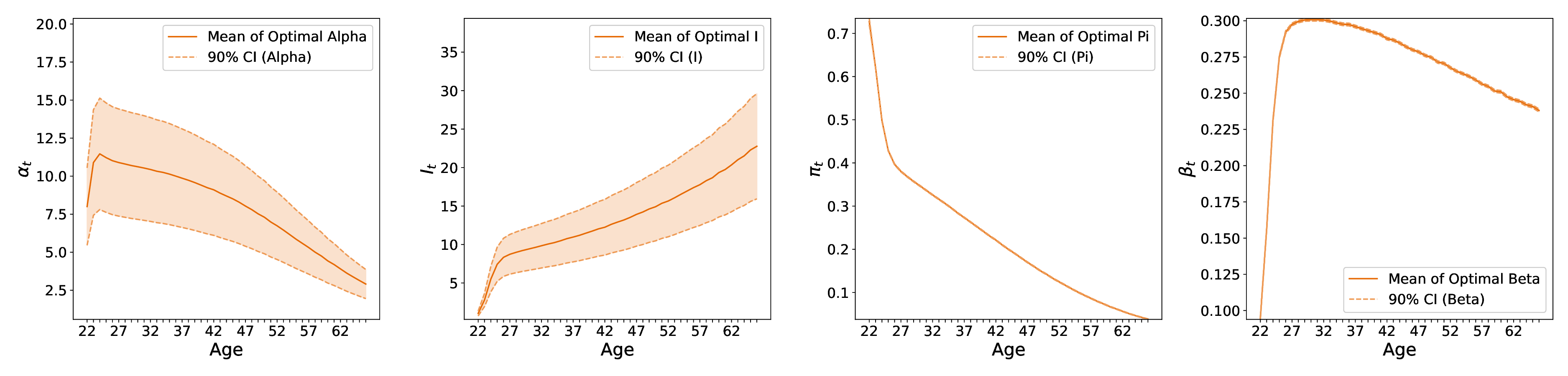}}

\vspace{-0.05in}
\subfigure[Non-periodic evaluation without mortality projection (2022)\label{fig:dynamic_np}]{
\includegraphics[width=\linewidth,height=0.28\textheight,keepaspectratio]{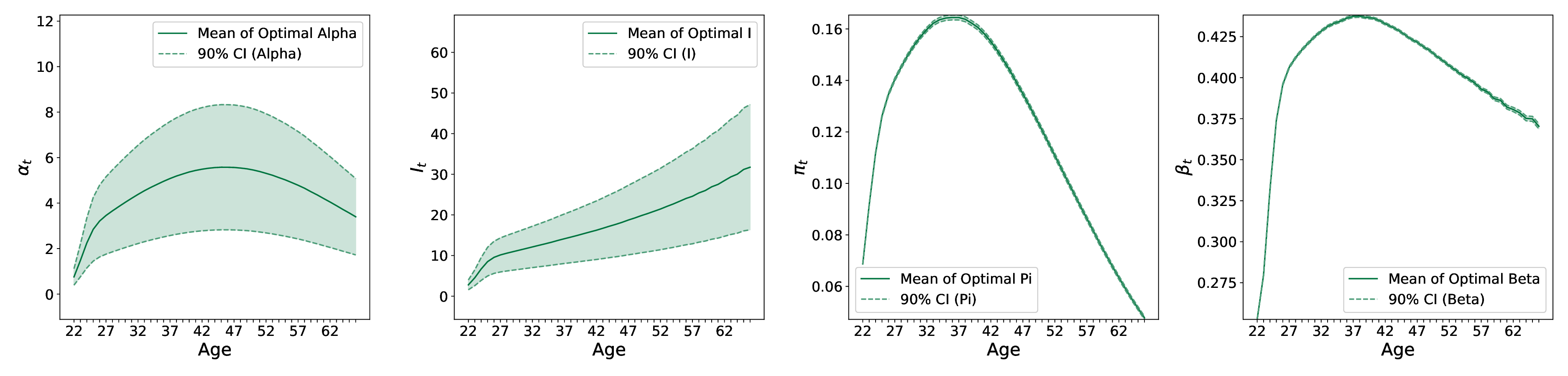}}

\caption{Optimal strategies $(\alpha^\ast_t,I^\ast_t)$ under mortality projection and periodic evaluation effects.}


\end{sidewaysfigure}

Figure \ref{fig:dynamic} and \ref{fig:dynamic_lc} present the dynamic optimal strategy mean $(\Bar{\alpha^\ast}, \Bar{I^\ast})$ over the working period for (a) the case without projections and (b) the case with Lee-Carter projections, based on 5,000 simulated trajectories of the state vectors $(t, W_t, Y_t)$ for $t=0,\ldots,44$. The shaded areas represent the $90$\% CIs of the marginal distributions for optimal allocations, highlighting the inherent uncertainty in dynamic optimal strategies over time. In each row, the four subfigures illustrate the intertemporal adjustments in: (i) $\alpha_t$ (dollar allocation to stocks, in thousands), (ii) $I_t$ (dollar allocation to term life insurance, in thousands), (iii) $\pi_t$ (proportion of total wealth allocated to risky investments, \%), and (iv) $\beta_t$ (proportion allocated to term life insurance, \%).

In the absence of mortality projections, we observe that the pension members become more risk-averse with age. Specifically, their dollar allocation to stocks ($\alpha_t$) gradually declines over time, while the allocation to term life insurance ($I_t$) increases steadily and peaks near retirement. This pattern aligns with the natural life cycle, where younger people focus on wealth accumulation while older people shift to risk mitigation. As individuals tend to live longer with mortality improvement, we observe that the pension members become more risk-seeking when considering the LC mortality projections. Specifically, they allocate more money into the risky asset in both absolute and proportional amounts, resulting in a larger accumulated wealth level (see Figure  \ref{fig:dynamic_lc} and $W_t$ column in Table \ref{tab:comparison}). From a proportional perspective, the allocation to stocks ($\pi_t$) is significantly higher in the Lee-Carter projection case, exceeding 20\% before age 45, compared to less than 10\% after age 25 in the no-projection case. The proportion allocated to life insurance ($\beta_t$) exhibits a hump-shaped pattern in both scenarios, peaking before age 27 in case (a) while around age 30 in case (b), and declining thereafter. This hump shape arises from the interplay between the wealth accumulation effect (positive) and the aging effect (negative). As wealth grows over time, members tend to pay more in insurance premiums, but the diminishing marginal gain from insurance and the decreasing time window reduce the optimal life insurance proportion in later years\footnote{Similar interrelationships have been discussed in the existing literature; see \cite{ameriks2011joy, jere2022optimal}.}. As a result, although the absolute amount of buying life insurance increases due to the increasing purchasing power(larger wealth level), the proportion of wealth purchasing life insurance decreases as pension members consider the mortality projection. In general, we conclude that when an individual lives longer, the amount of money allocated to life insurance increases with larger purchasing power, while the proportion of wealth allocated to life insurance decreases due to the shrinking demand.

\begin{table}[htbp]
    \caption{Comparison of optimal strategies under different mortality assumptions.}
    \smallskip
    \centering
    \begin{tabular}{ccccccccc}
    \hline
    \multicolumn{8}{c}{\textbf{Without Mortality Projection}} \\
    Timestep & Age &$\Bar{W}_t$ & $\Bar{Y}_t$ & $\mu_\alpha$ & $\mu_I$ & $\sigma_\alpha$ & $\sigma_I$ & $\rho_{\alpha,I}$ \\
    \hline
    0 & 22 & 5 & 60 & 1.4390 & 5.5735 & 0.22778 & 0.0018 & -0.7700 \\
    22 & 44 & 12.7128 & 119.4110 & 0.0596 & 11.7134 & 0.0110 & 0.0010 & -0.9621 \\
    44 & 66 & 25.7242 & 237.7540 & 0.0003 & 22.1453 & 0.0010 & 0.0010 & -0.9983 \\
    \hline
    \multicolumn{8}{c}{\textbf{With Lee-Carter Projection}} \\
    Timestep & Age & $\Bar{W}_t$ & $\Bar{Y}_t$ & $\mu_\alpha$ & $\mu_I$ & $\sigma_\alpha$ & $\sigma_I$ & $\rho_{\alpha,I}$ \\
    \hline
    0 & 22 & 5 & 60 & 8.0001 & 1.0430 & 1.5462 & 0.1945 & -0.9930 \\
    22 & 44 & 33.6093 & 118.8920 & 8.9120 & 13.1132 & 1.7820 & 2.3995 & -0.9990 \\
    44 & 66 & 73.5079 & 236.6450 & 2.2766 & 25.5023 & 0.4553 & 4.9540 & -0.9990 \\
    \hline
    \end{tabular}
    \label{tab:comparison}
\end{table}

\subsection{Periodic evaluation effect}\label{subsec_nonperiodic}

This section studies the periodic evaluation effects on the DC pension management. The random costs follow equation \eqref{pe_cost} for the periodic evaluation and equation \eqref{np_cost} for the non-periodic evaluation. The numerical results are shown in Table \ref{tab_pvsnp} and Figure \ref{fig:dynamic_np}.

\vspace{-0.2cm}
\begin{table}[ht]
    \caption{Comparison of optimal strategies by periodic vs. non-periodic evaluations.}
    \smallskip
    \centering
    \begin{tabular}{c|cccccc}
        \hline
        \textbf{Evaluation Type} & Timestep & $W_t$ & $\alpha_t$ & $I_t$ & $\pi_t$ (\%) & $\beta_t$ (\%) \\
        \hline
        \multirow{3}{*}{\textbf{Periodic}} 
        & 0 & 5.0000 & 1.4390 & 5.5735 & 13.08 & 50.67 \\
        & 22 & 12.7128 & 0.0598 & 11.7273 & 0.26 & 47.10 \\
        & 44 & 25.7242 & 0.0004 & 23.7887 & 8.54$\times 10^{-4}$ & 44.48 \\
        \hline
        \multirow{3}{*}{\textbf{Non-periodic}} 
        & 0 & 5.0000 & 0.7563 & 2.7791 & 6.88 & 25.26 \\
        & 22 & 28.4623 & 5.5613 & 17.1967 & 14.78 & 42.90 \\
        & 44 & 61.8531 & 3.3983 & 31.7273 & 4.78 & 37.04 \\
        \hline
    \end{tabular}
    \label{tab_pvsnp}
\end{table}
In general, evaluating annually can be viewed as imposing an additional trading constraint on the non-periodic evaluator. Since the non-periodic evaluator does not require annual evaluation, they have the flexibility to choose whether to be evaluated annually or not. Consequently, the non-periodic evaluator's admissible set of trading strategies is larger than that of the periodic evaluator. Table \ref{tab_pvsnp} presents a comparative analysis between the periodic and non-periodic evaluators. The results indicate that the periodic evaluator tends to reduce their allocation to the risky asset over time, both in absolute and proportional terms. Conversely, their absolute allocation to life insurance increases as time progresses, while the proportion allocated to life insurance follows a hump-shaped pattern. The non-periodic evaluator exhibits a more risk-seeking behavior; they tend to purchase more stocks during middle age and decrease their holdings as they approach retirement. As a result, the non-periodic evaluator generally attains a higher wealth level compared to the periodic evaluator. Additionally, their absolute demand for life insurance exceeds that of the periodic evaluator, attributable to their higher wealth. In contrast, their proportional demand for life insurance is lower, reflecting their risk-seeking tendencies.

Figure \ref{fig:dynamic_np} shows the optimal investment and insurance strategies for the non-periodic evaluator. Compared to the periodic evaluator (see Figure \ref{fig:dynamic}), the non-periodic evaluator exhibits a hump-shaped pattern in stock investment, both in absolute terms and proportionally. As there is no annual evaluation pressure for the non-periodic evaluator, they can make bolder investments in their middle age. Consistent with the results in Table \ref{tab_pvsnp}, the non-periodic evaluator allocates a larger absolute amount of wealth to life insurance due to their higher overall wealth level. However, their proportional allocation to life insurance remains low, reflecting their risk-seeking preferences.
 
\begin{figure}[htbp]
    \centering
    \includegraphics[width=0.6\textwidth]{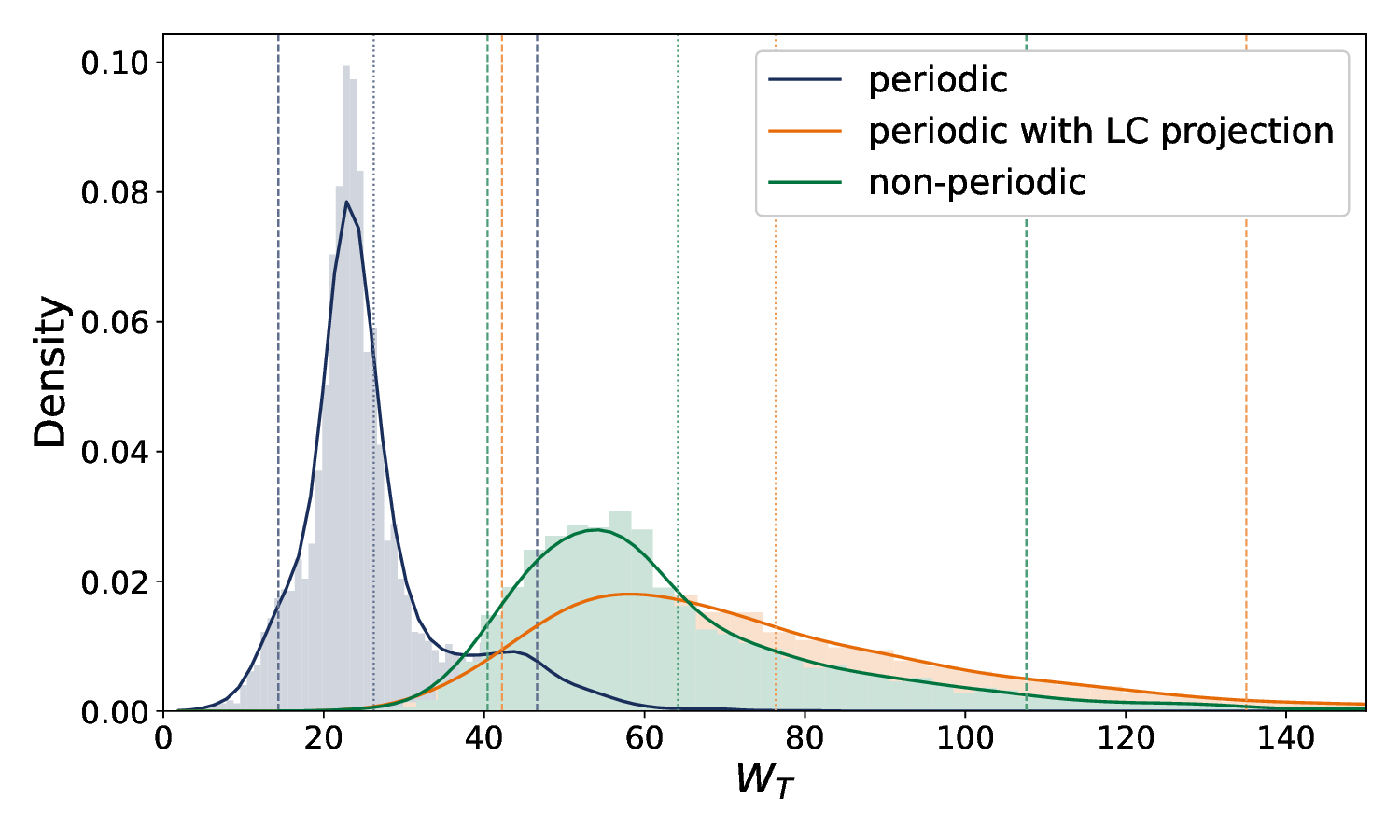}   
    \vspace{-0.5cm}
    \caption{Distributions of terminal wealth by periodic vs. non-periodic evaluations. The vertical dashed lines are the $5\%$, $50\%$, and $95\%$ quantiles of the corresponding distributions.}
    \label{fig:xT}
\end{figure}

Lastly, we also plot the distribution of the terminal wealth for each evaluator in Figure \ref{fig:xT}. We see that the periodic evaluator earns a larger terminal wealth when considering the LC mortality projection. This is a consequence of the risk-seeking investment coming from the optimist about their future life. For the non-periodic evaluator, the distribution of their terminal wealth is flatter and right-tail distributed than that of the periodic evaluator. This is caused by the aggressive investment-insurance strategy, as the non-periodic evaluator doesn't have the annual evaluation pressure. Comparisons among these three evaluators illustrate the essentials of mortality projection and evaluation frequency on DC pension management.

\section{Conclusion}\label{sec_conclusion}
This study proposes a novel framework for the periodic evaluation of DC pension funds using the dynamic risk measure. The pension member allocates funds among stocks, bonds, and life insurance, making joint decisions with the beneficiary. We develop a dynamic programming approach for these decision-makers and employ the actor-critic algorithm to determine optimal strategies. Our numerical analysis compares three evaluators: the periodic evaluator without mortality projection, the periodic evaluator with mortality projection, and the non-periodic evaluator without mortality projection. The results indicate that both the mortality projection and the non-periodic evaluation encourage more risk-seeking behaviors, such as increased stock allocations and reduced life insurance proportions, leading to higher terminal wealth at retirement. Based on these findings, we recommend that DC pension members adopt a periodic evaluation framework to stabilize their retirement portfolios and incorporate mortality projections to further enhance investment outcomes. Lastly, future research could extend this framework by integrating additional risk factors and advanced reinforcement learning algorithms, thus improving pension members’ ability to navigate financial uncertainties.

\section*{Acknowledgements}
Wenyuan Li gratefully acknowledges a start-up grant from the University of Hong Kong. Yunran Wei acknowledges financial support from the Natural Sciences and Engineering Research Council of Canada (RGPIN-2023-04674, DGECR-2023-00454), the Research Development Grants, and the start-up fund at Carleton University.

\bibliographystyle{apalike}
\bibliography{reference}

\appendix
\section{Proof of Theorem \ref{approxV_thm}}\label{appendA}
\begin{proof}
    The value function $V_t$ may be written recursively with the DPE/recursive formula (\ref{dpe1})
    $$ V_t = q_{x+t}\rho_t^B(c_t^B)+p_{x+t}\rho_t^A(c_t^A+e^{-r}V_{t+1}).$$
    We show that the ANN $V^{\phi}_t$ converges to the true value function by induction. 
    At $t=T$, the terminal value function is given by $$ V_T = \rho^A_{T}\left(- W^{\theta}_{T} \right).$$
    Since $\rho_t^A$  is a convex risk measure, which is Lipschitz continuous, by the universal approximation theorem from \cite{cybenko1989approximation}, there exists an ANN $V^{\phi}_T$ such that for any $\epsilon_{T},$
    $$\underset{s \in S}{\operatorname{ess} \sup }\left\|V_{T}(s ; \theta)-V_{T}^{\phi}(s ; \theta)\right\|<\epsilon_{T}.$$
    Suppose that the statement holds at $t+1$, that is, for any $\epsilon_{t+1}$, there exists an ANN $V_{t+1}^\phi$ that approximates $V_{t+1}$
    $$\underset{s \in S}{\operatorname{ess} \sup }\left\|V_{t+1}(s ; \theta)-V_{t+1}^{\phi}(s ; \theta)\right\|<\epsilon_{t+1}.$$
    Plug the ANN $V_{t+1}^\phi$ into the DPE \eqref{dpe1}, we obtain the intermediate function $\tilde{V}_t$
    $$\tilde{V}_t= q_{x+t}\rho_t^B(c_t^B)+p_{x+t}\rho_t^A(c_t^A+e^{-r}V_{t+1}^\phi).$$
    Recall that both $\rho_t^A$ and $\rho_t^B$ are convex risk measures, and $\tilde{V}_t$ is absolutely continuous a.s. By the universal approximation theorem, there exists an ANN, denoted by $V_t^\phi$, such that 
    $$ \underset{s \in S}{\operatorname{ess} \sup }\left\|\tilde{V}_{t}(s ; \theta)-V_{t}^{\phi}(s ; \theta)\right\|<\epsilon_{t}.$$
    Next, at time $t$, we have
    \begin{align*}
     &   \underset{s \in \mathcal S}{\operatorname{ess} \sup }\left\|V_{t}(s ; \theta)-V_{t}^{\phi}(s ; \theta)\right\| \\     \leq &\operatorname{ess} \sup _{s \in \mathcal{S}}\left\|V_t(s ; \theta)-\tilde{V}_t(s ; \theta)\right\|+\operatorname{ess} \sup _{s \in \mathcal{S}}\left\|\tilde{V}_t(s ; \theta)-V_t^\phi(s ; \theta)\right\|\\
    < & \operatorname{ess} \sup _{s \in \mathcal{S}}\Big\|q_{x+t}\rho_t^B(c_t^B)+p_{x+t}\rho_t^A(c_t^A+e^{-r}V_{t+1}(s;\theta) )- (q_{x+t}\rho_t^B(c_t^B)\\
     & \qquad +p_{x+t}\rho_t^A(c_t^A+e^{-r}V_{t+1}^\phi(s;\theta))) \Big\| + \epsilon_t \\
     \leq & k \operatorname{ess} \sup _{s \in \mathcal{S}}||e^{-r}(V_{t+1}(s;\theta) - V^\phi_{t+1}(s;\theta))||+\epsilon_t\\
     \le & ke^{-r}\epsilon_{t+1} + \epsilon_t,
    \end{align*}
    where the second last inequality follows from the Lipschitz continuous property of $\rho_t^A$ and $k$ is a non-negative constant. 
\end{proof}

\section{Proof of Theorem \ref{gradV_thm}}\label{appendB}
\begin{proof}
    With the recursive formulation of dynamic risk measures, we can compute the value function gradient from the last term and then extend it for the subsequent periods. Analogous to the DPE \eqref{dpe1} for the value function $V_t$, we have the same recursive formulation for the gradient of $V_t$. Without loss of generality, we only show the last-term gradient. By Definition \ref{riskenvelope} and equation \eqref{minimax}, for $t=T-1$, the Lagrangian of the maximization problem for any state $s\in\S$ can be written as
    \begin{footnotesize}
    \begin{align*}
    &\L^{\theta}(\bm{\xi},\bm{\lambda},\lambda^{\mathcal{E}},\lambda^{\mathcal{I}}) = p_{x+T-1}\left(\sum_{(a,s^\prime)}\xi^A(a,s^\prime)\P^{\theta}(a,s^{\prime}\mid s_{T-1}=s)\left(c^A_{T-1}(s,a,s^{\prime}) +e^{-r}V_T(s^{\prime};\theta)\right)- \rho^{A,\ast}_{T-1}(\xi^A)\right) \\
    & + q_{x+T-1}\left(\sum_{(a,s^\prime)}\xi^B(a,s^\prime)\P^{\theta}(a,s^{\prime}\mid s_{T-1}=s)c^B_{T-1}(s,a,s^{\prime}) - \rho^{B,\ast}_{T-1}(\xi^B)\right)\\
    & -\lambda^A p_{x+T-1}\Big(\sum_{(a,s^\prime)}\xi^A(a,s^\prime)\P^{\theta}(a,s^\prime \mid s_{T-1}=s)-1\Big) -\lambda^B q_{x+T-1}\Big(\sum_{(a,s^\prime)}\xi^B(a,s^\prime)\P^{\theta}(a,s^\prime \mid s_{T-1}=s)-1\Big)\\
    & - \sum_{e\in\mathcal{E}}\lambda^{\mathcal{E}}(e)\left(p_{x+T-1}g_e(\xi^A,\P^{\theta}) + q_{x+T-1}g_e(\xi^B,\P^{\theta})\right) - \sum_{i\in\mathcal{I}}\lambda^{\mathcal{I}}(i)\left(p_{x+T-1}f_i(\xi^A,\P^{\theta}) + q_{x+T-1}f_i(\xi^B,\P^{\theta})\right).
    \end{align*}
    \end{footnotesize}
    According to the convexity of \eqref{dpe2} and Definition \ref{riskenvelope}, $\L^{\theta}$ has at least one saddle-point. We emphasize that the saddle points $\left(\bm{\xi}^\ast, \bm{\lambda}^\ast, \lambda^{\ast, \mathcal{E}}, \lambda^{\ast, \mathcal{I}}\right)$ depend on the state $s$. By Slater's condition (\citeauthor{slater2013lagrange}, \citeyear{slater2013lagrange}), the strong duality holds, such that 
    $$
    V_{T-1}(s;\theta) = \max_{\bm{\xi}\geq 0}\min_{\bm{\lambda},\lambda^{\mathcal{E}},\lambda^{\mathcal{I}}}\L^{\theta}(\bm{\xi}, \bm{\lambda},\lambda^{\mathcal{E}},\lambda^{\mathcal{I}}).
    $$
    Now we apply it to solve for the saddle-point problems following the Envelope theorem (\citeauthor{milgrom2002envelope}, \citeyear{milgrom2002envelope}). Since our objective function is equidifferentiable, and the gradient of the value function is absolutely continuous by Assumption \ref{assump1}, we have 
    $$
    \max_{\bm{\xi}\geq 0}\min_{\bm{\lambda},\lambda^{\mathcal{E}},\lambda^{\mathcal{I}}}\L^{\theta}(\bm{\xi}, \bm{\lambda},\lambda^{\mathcal{E}},\lambda^{\mathcal{I}}) = \max_{\bm{\xi}\geq 0}\min_{\bm{\lambda},\lambda^{\mathcal{E}},\lambda^{\mathcal{I}}}\L^{0}(\bm{\xi}, \bm{\lambda},\lambda^{\mathcal{E}},\lambda^{\mathcal{I}}) + \int^{\theta}_0\nabla_{\theta}\L^{\theta}(\bm{\xi}, \bm{\lambda},\lambda^{\mathcal{E}},\lambda^{\mathcal{I}})\bigg|_{\theta=u}\mathrm{d}u,
    $$
    which provides equality between $\nabla_{\theta}V_{T-1}$ and $\nabla_{\theta}\L^{\theta}$ at any saddle point. As a result, we can obtain the following gradient of the value function by the likelihood ratio trick in \eqref{probtrick}
    \begin{small}
    \begin{align*}
        &\nabla_{\theta}V_{T-1}(s;\theta) =  \nabla_{\theta} \max_{\bm{\xi}\geq 0}\min_{\bm{\lambda},\lambda^{\mathcal{E}},\lambda^{\mathcal{I}}}\L^{\theta}(\bm{\xi}, \bm{\lambda},\lambda^{\mathcal{E}},\lambda^{\mathcal{I}})=  \nabla_{\theta}\L^{\theta}(\bm{\xi}, \bm{\lambda},\lambda^{\mathcal{E}},\lambda^{\mathcal{I}})\Big|_{\bm{\xi}^\ast, \bm{\lambda}^\ast, \lambda^{\ast, \mathcal{E}}, \lambda^{\ast, \mathcal{I}}}\\
        & =  p_{x+T-1} \left(\sum_{(a,s^\prime)}\xi^{A,\ast}(a,s^\prime)\nabla_{\theta}\P^{\theta}(a,s^{\prime}\mid s)\left(c^A_{T-1}(s,a,s^{\prime}) + e^{-r} V_T(s^\prime;\theta)\right) \right.\\
        &\qquad + \left.\sum_{(a,s^\prime)}\xi^{A,\ast}(a,s^\prime)\P^{\theta}(a,s^\prime\mid s)\nabla_{\theta}V_T(s^\prime;\theta) - \nabla_{\theta}\rho^{A,\ast}_{T-1}(\xi^A)\right) \\
        & + q_{x+T-1} \left(\sum_{(a,s^\prime)}\xi^{B,\ast}(a,s^\prime)\nabla_{\theta}\P^{\theta}(a,s^{\prime}\mid s)c^B_{T-1}(s,a,s^{\prime}) - \nabla_{\theta}\rho^{B,\ast}_{T-1}(\xi^B)\right)\\
        & -\lambda^{A,\ast} p_{x+T-1} \sum_{(a,s^\prime)}\xi^{A,\ast}(a,s^\prime)\nabla_{\theta}\P^{\theta}(a,s^\prime \mid s) -\lambda^{B,\ast} q_{x+T-1} \sum_{(a,s^\prime)}\xi^{B,\ast}(a,s^\prime)\nabla_{\theta}\P^{\theta}(a,s^\prime \mid s)\\
        & - \sum_{e\in\mathcal{E}}\lambda^{\mathcal{E},\ast}(e)\left(p_{x+T-1}\nabla_{\theta}g_e(\xi^{A,\ast},\P^{\theta}) + q_{x+T-1}\nabla_{\theta}g_e(\xi^{B,\ast},\P^{\theta})\right)\\
        & - \sum_{i\in\mathcal{I}}\lambda^{\mathcal{I},\ast}(i)\left(p_{x+T-1}\nabla_{\theta}f_i(\xi^{A,\ast},\P^{\theta}) + q_{x+T-1}\nabla_{\theta}f_i(\xi^{B,\ast},\P^{\theta})\right) \\
        &= p_{x+T-1}\E^{\xi^{A,\ast}}_{T-1}\bigg[ (c^A_{T-1}(s,a^{\theta}_{T-1},s^{\theta}_T) + e^{-r}V_T(s^{\theta}_{T};\theta)-\lambda^{A,\ast}) \nabla_{\theta}\log\pi^{\theta}(a^{\theta}_{T-1}\mid s) + \nabla_{\theta}V_T(s^{\theta}_T;\theta)\bigg]\\
        &+ q_{x+T-1}\E^{\xi^{B,\ast}}_{T-1}\bigg[ (c^B_{T-1}(s,a^{\theta}_{T-1},s^{\theta}_T) -\lambda^{B,\ast}) \nabla_{\theta}\log\pi^{\theta}(a^{\theta}_{T-1}\mid s) \bigg] \\
        & -p_{x+T-1}\nabla_{\theta}\rho^{A,\ast}_{T-1}(\xi^{A,\ast})- q_{x+T-1}\nabla_{\theta}\rho^{B,\ast}_{T-1}(\xi^{B,\ast}) \\
        & - \sum_{e\in\mathcal{E}}\lambda^{\mathcal{E},\ast}(e)\left(p_{x+T-1}\nabla_{\theta}g_e(\xi^{A,\ast},\P^{\theta}) + q_{x+T-1}\nabla_{\theta}g_e(\xi^{B,\ast},\P^{\theta})\right)\\
        & - \sum_{i\in\mathcal{I}}\lambda^{\mathcal{I},\ast}(i)\left(p_{x+T-1}\nabla_{\theta}f_i(\xi^{A,\ast},\P^{\theta}) + q_{x+T-1}\nabla_{\theta}f_i(\xi^{B,\ast},\P^{\theta})\right).
    \end{align*}
    \end{small}
    This concludes the proof.
\end{proof}

\end{document}